\documentclass[prd,onecolumn,nofootinbib]{revtex4}
\usepackage{bm,amsmath,amssymb,graphicx}
\begin{document}

\title{Gravitational field of spinors with torsion and matter--antimatter asymmetry}
\author{Nikodem Pop{\l}awski}

\altaffiliation{NPoplawski@newhaven.edu}
\affiliation{Department of Mathematics and Physics, University of New Haven, West Haven, CT, USA}

\begin{abstract}
The conservation law for the orbital plus spin angular momentum of a free Dirac particle in curved spacetime requires that the affine connection has the antisymmetric part: the torsion tensor, which extends general relativity to the Einstein--Cartan theory of gravity.
In spacetime with torsion, the Dirac equation becomes a nonlinear, cubic equation in the spinor wave function.
Under charge conjugation, the nonlinear term changes sign relative to the mass term.
A Dirac spinor and its charge conjugate therefore satisfy different field equations.
We use the spinor connection formalism to derive the Dirac equation for a spherical spinor in its centrally symmetric gravitational field with torsion.
We determine the energy--momentum tensor for this spinor and show that the energy contribution from torsion is negative for matter particles and positive for antimatter particles.
Consequently, the ground state of a fermion has a lower energy than the ground state of the corresponding antifermion.
We also derive the Einstein--Cartan--Dirac equations for this spinor; they indicate the existence of particle-like soliton solutions exhibiting matter--antimatter energy asymmetry.
\end{abstract}
\maketitle

{\bf 1. Introduction}\\ \\
The Dirac equation, formulated by Paul Dirac in 1928, successfully merged quantum mechanics with special relativity \cite{Dirac}.
This relativistic wave equation naturally explained the origin of the intrinsic spin angular momentum of the electron and explained the fine structure of the hydrogen atom.
To reconcile the resulting negative energy eigenvalues with atomic stability, Dirac proposed in 1931 the hole theory, conceptualizing the quantum vacuum as a fully occupied infinite sea of negative energy states, in which an unoccupied vacancy manifests itself as an antiparticle (positron) with the same mass as the particle (electron) but with the opposite electric charge \cite{Diracpositron}.
This theoretical prediction of antimatter was verified in 1932 by Anderson, who discovered the positron in cosmic rays \cite{Anderson}.

While the ALPHA-g experiment conducted in 2023 at CERN directly confirmed that antimatter obeys the weak equivalence principle by falling downward under gravity identically to ordinary matter \cite{alphag}, this symmetry deepens a foundational cosmological paradox.
Standard big bang cosmology dictates that matter and antimatter should have been created in equal amounts and completely annihilated each other in the early Universe.
Why our Universe is comprised entirely of matter remains one of the greatest mysteries in modern science.

Three necessary criteria that a physical process must satisfy for the Universe to develop an imbalance between matter and antimatter were proposed in 1967 by Sakharov: baryon number violation, charge conjugation (C) symmetry violation and charge conjugation plus parity (CP) symmetry violation, and a departure from thermal equilibrium \cite{Sakharov}.
CP violation has been verified in decays of neutral mesons: kaons, B-mesons, and D-mesons, and beauty baryons \cite{decays}.
Contemporary proposals to explain the missing antimatter range from including baryogenesis and leptogenesis from CP violation \cite{review}, through asymmetry in gravitational capture by black holes \cite{capture}, to the hypothesis that vast, undetected antimatter galaxies exist in remote sectors of the observable Universe, which was ruled out in \cite{antigalaxy}.

In this article, we show that spacetime torsion \cite{Schr} naturally violates C-symmetry of fermions and generates the matter--antimatter asymmetry.
We use the Einstein--Cartan--Sciama--Kibble theory of gravity \cite{EC} which naturally extends general relativity to include the intrinsic spin of matter \cite{Lord,FranciscoNiko}.
Spacetime torsion produced by spin generates in the Dirac equation a term which is cubic in spinor fields, extending it into the Hehl--Datta equation \cite{HehlDatta}.
We showed that under the charge-conjugation transformation, this nonlinear term changes sign relative to the mass term \cite{anti}.
A Dirac spinor and its charge conjugate therefore satisfy different field equations.

We consider the Wakano ansatz for a spherical spinor \cite{Wakano,RQM}.
This spinor coupled to the Maxwell equations in flat spacetime can produce stable localized particle-like solutions (solitons) to represent the electron \cite{Wakano}.
The solitary wave is formed by the nonlinearity in the Dirac equation arising from the electromagnetic potential being sourced by the Dirac current.
Stable localized configurations with a lowest energy state can also exist for a spinor field in flat spacetime with the simplest nonlinear self-interaction, proposed by Soler \cite{Soler}.
We use the spin connection and the Fock--Ivanenko spinor connection \cite{FockIvanenko} to derive the Dirac equation for a fermion in its static centrally symmetric gravitational field with torsion and analyze how it changes under charge conjugation, extending our earlier work \cite{anti}.
Without torsion, this equation and the Einstein equations were derived for a static, spherically symmetric system of two fermions in a singlet spinor state, showing the existence of particle-like soliton solutions \cite{Smoller}.
Instead of spinor connection, that derivation used the Newman--Penrose formalism, which was calculated earlier by Chandrasekhar \cite{Chandra} to prove that the Dirac equation can be separated into the radial and angular parts for the Kerr metric describing a rotating black hole \cite{Kerr}.
The derivation from the spin connection \cite{rotations} is more straightforward for conceptual, structural calculations and shows that a specific unitary rotation of a spinor can eliminate two angular spinor connections from the Dirac equation, simplifying the separation of the radial and angular parts.

We show that a free antifermion at rest, represented by the spherical spinor solution of the Dirac equation has a higher energy than the corresponding fermion at rest, caused by the nonlinear contribution from torsion.
This contribution is analogous to the nonlinear corrections considered in \cite{Wakano} and \cite{Soler}.
This energy difference, proposed in \cite{anti}, might have led to the matter--antimatter imbalance in the early Universe.
A numerical analysis of the Einstein-hydrodynamics equations describing ultrarelativistic head-on collisions of fluid particles shows that increasing the energy of particles lowers the threshold for black hole formation \cite{numerical}.
Consequently, antimatter particles might have been more likely to form microscopic primordial black holes \cite{primordial}, which later might have become supermassive black holes at the centers of galaxies.
This mechanism could explain the matter--antimatter imbalance in the observable Universe while preserving the total baryon number: the missing antimatter would be in black holes.

This article is structured as follows.
Section 2 explains torsion.
Section 3 summarizes the Einstein--Cartan theory of gravity.
Section 4 summarizes spinors and the covariant Dirac equation.
Section 5 shows the derivation of the spin tensor and how torsion makes the Dirac equation nonlinear in the spinor field.
Section 6 shows the derivation of the energy--momentum tensor for the spinor field.
In Section 7, we use the spin connection and the Fock--Ivanenko spinor connection to derive the Dirac equation in a centrally symmetric gravitational field.
Section 8 shows how to apply a unitary rotation of a spinor to eliminate two angular spinor connections from the Dirac equation to simplify the separation of the radial and angular parts.
In Section 9, we derive how the curvature and torsion terms in the Dirac equation transform under this rotation.
In Section 10, we show how a spherical spinor satisfies the Dirac equation with curvature and torsion, calculating the cubic term in the Dirac equation and obtaining nonlinear radial equations for the spinor wave function.
In Section 11, we show how charge conjugation changes the relative sign between the mass term and the cubic term in the Dirac equation.
We also show this asymmetry explicitly for a spherical spinor.
In Section 12, we calculate the energy--momentum tensor for a spherical spinor with curvature and torsion.
We show that the energy contribution from torsion is negative for matter particles and positive for antimatter particles.
Consequently, the ground state of a
fermion has a lower energy than the ground state of the corresponding antifermion. 
In Section 13, we derive the Einstein--Cartan--Dirac equations for a spherical spinor in its static, centrally symmetric gravitational field and conclude the article.\\

{\bf 2. Torsion}\\ \\
The affine connection $\Gamma^k_{ij}$ is used to construct the covariant derivative $\nabla_i$ of a vector that transforms under coordinate transformations like a tensor:
\[
    \nabla_i V^k=\partial_i V^k+\Gamma^k_{ji}V^j,\quad \nabla_i V_k=\partial_i V_k-\Gamma^j_{ki}V_j.
\]
For tensors with more indices, each index produces a term with the affine connection.
The affine connection is not a tensor, but its variation is a tensor.
The antisymmetric part of the affine connection is also a tensor, called the torsion tensor \cite{Schr,EC,Lord,FranciscoNiko}:
\begin{equation}
    S^k_{\phantom{k}ij}=\frac{1}{2}(\Gamma^{k}_{ij}-\Gamma^{k}_{ji}).
    \label{torsion}
\end{equation}

The metric tensor $g_{ik}$ is used to define a distance in a curved coordinate system:
$ds^2=g_{ik}dx^i dx^k$, where $s$ is the interval.
The inverse metric tensor $g^{ik}$ satisfies $g_{ik}g^{jk}=\delta^j_i$, where $\delta^j_i$ is the Kronecker tensor.
Both metric tensors relate the contravariant and covariant vector indices: $V_i=g_{ik}V^k,\,V^i=g^{ik}V_k$.

The covariant derivative of the metric tensor is zero, which is called metricity: $\nabla_j g_{ik}=0$.
Therefore, the covariant differentiation commutes with converting between contravariant and covariant indices.
This relation determines the affine connection in terms of the metric tensor, its partial derivatives, and the torsion tensor \cite{EC,Lord,FranciscoNiko}:
\[
    \Gamma^{m}_{ik}=\mathring{\Gamma}^{m}_{ik}+S^m_{\phantom{m}ik}+S_{ik}^{\phantom{ik}m}+S_{ki}^{\phantom{ki}m},
\]
where the Christoffel symbols are given by
\begin{equation}
    \mathring{\Gamma}^{m}_{ik}=\frac{1}{2}g^{ml}(\partial_i g_{kl}+\partial_k g_{il}-\partial_l g_{ik}).
    \label{affineconnection}
\end{equation}
They form a connection, called the Levi-Civita connection.
The difference between the two connections is the contortion tensor:
\begin{equation}
    C^k_{\phantom{k}ij}=S^k_{\phantom{k}ij}+S_{ij}^{\phantom{ij}k}+S_{ji}^{\phantom{ji}k}.
    \label{contortion}
\end{equation}

The curvature tensor is given by
\[
    R^i_{\phantom{i}klm}=\partial_{l}\Gamma^{i}_{km}-\partial_{m}\Gamma^{i}_{kl}+\Gamma^{j}_{km}\Gamma^{i}_{jl}-\Gamma^{j}_{kl}\Gamma^{i}_{jm}.
\]
This tensor can be decomposed into the Riemann tensor $\mathring{R}^i_{\phantom{i}klm}$, which is constructed from the $\mathring{\Gamma}^{j}_{ik}$ in the same way as the curvature tensor is constructed from $\Gamma^{j}_{ik}$, and a part that includes the contortion tensor:
\[
    R^i_{\phantom{i}klm}=\mathring{R}^i_{\phantom{i}klm}+\mathring{\nabla}_{l}C^i_{\phantom{i}km}-\mathring{\nabla}_{m}C^i_{\phantom{i}kl}+C^j_{\phantom{j}km}C^i_{\phantom{i}jl}-C^j_{\phantom{j}kl}C^i_{\phantom{i}jm},
\]
where $\mathring{\nabla}_{i}$ is the covariant derivative with respect to the Levi-Civita connection.

Torsion has a simple geometrical interpretation.
In the presence of curvature, a parallel translation (which also defines the covariant derivative) of a vector around a closed curve changes the vector.
In the presence of torsion, such a translation does not change the vector, but two parallel translations do not commute, which results from the following construction \cite{Schr,FranciscoNiko}.
The parallel translation of an infinitesimal, four-dimensional vector $PR=dx^i$ from a point $P$ to an infinitesimally close point $Q$ such that $PQ=dx'^j$ adds to $dx^i$ a small correction: $\delta dx^i = -\Gamma^{i}_{jk}dx^j dx'^k$ \cite{LL2}.
After effecting the translation, the vector $dx^i+\delta dx^i$ points to a point $T$.
The parallel translation of the vector $dx'^i$ from a point $P$ to an infinitesimally close point $R$ adds to $dx^i$ a small correction: $\delta dx'^i = -\Gamma^{i}_{jk}dx^j dx'^k$.
After effecting the translation, the vector $dx'^i+\delta dx'^i$ points to a point $T'$.

\begin{figure}[th]
    \centering
    \includegraphics[width=1.2in]{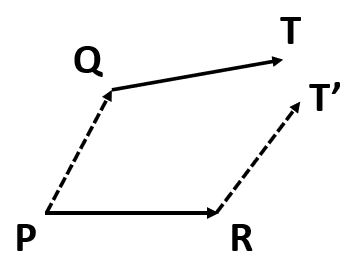}
    \caption{Nonclosure of the parallelogram formed from two vectors translated along each other depends on torsion.}
    \label{parallel}
\end{figure}

Without torsion, the points $T$ and $T'$ would coincide and form, together with the points $P$, $Q$, and $R$, a parallelogram because $\delta dx'^i-\delta dx^i=\Gamma^{i}_{kj}dx^j dx'^k - \Gamma^{i}_{jk}dx^j dx'^k = 0$.
However, if the torsion tensor is not zero, then the affine connection is asymmetric in the lower indices and $\delta dx'^i-\delta dx^i = -2S^i_{\phantom{i}jk}dx^j dx'^k$.
The points $T$ and $T'$ do not coincide, the distance between them is given by the torsion, the parallelogram is not closed, and the combination of two displacements of the point $P$ (through $dx^i$ and $dx'^j$) depends on their order, as shown in Figure \ref{parallel}.
Consequently, the covariant derivatives of a scalar $\psi$ do not commute: $\nabla_i\nabla_j\psi-\nabla_j\nabla_i\psi=\partial_i\nabla_j\psi-\Gamma^{k}_{ji}\nabla_k\psi-\partial_j\nabla_i\psi+\Gamma^{k}_{ij}\nabla_k\psi=\partial_i\partial_j\psi-\partial_j\partial_i\psi+2S^k_{\phantom{k}ij}\nabla_k\psi=2S^k_{\phantom{k}ij}\nabla_k\psi$.

At every point in spacetime, in addition to a general coordinate system, it is possible to set up four linearly independent vectors $e^i_\mu$ such that $e^i_\mu e^k_\nu g_{ik}=\eta_{\mu\nu}$ \cite{EC,Lord,FranciscoNiko}. 
This set of four vectors is called a tetrad and is used to construct the locally flat geometry.
The tetrad relates the Greek vector indices of the locally flat, Lorentz coordinate system (of the special theory of relativity) to the Latin vector indices of the general coordinate system (of general relativity): $V^i=e^i_\mu V^\mu$.
The choice of tetrad is not unique: a Lorentz transformation $\tilde{e}^i_\mu=\Lambda^\nu_\mu e^i_\nu$, where $\Lambda^\nu_\mu$ are the Lorentz matrices satisfying $\Lambda^\rho_\mu \Lambda^\sigma_\nu\eta_{\rho\sigma}=\eta_{\mu\nu}$, produces a new tetrad $\tilde{e}^i_\mu$.
The inverse tetrad $e_i^\mu$ satisfies $e^i_\mu e_i^\nu=\delta^\nu_\mu$ and $e^i_\mu e_k^\mu=\delta^i_k$.

The spin connection \cite{EC,Lord,FranciscoNiko}
\begin{equation}
    \omega^\mu_{\phantom{\mu}\nu i}=e^\mu_k\omega^k_{\phantom{k}\nu i}=e^\mu_k \nabla_i e^k_\nu=e^\mu_k(\partial_i e^k_\nu+\Gamma^{k}_{ji}e^j_\nu).
    \label{spinconnectiontetrad}
\end{equation}
extends covariant differentiation to vectors with Lorentz indices:
\[
    \nabla_i V^\mu=\partial_i V^\mu+\omega^\mu_{\phantom{\mu}\nu i}V^\nu,\quad \nabla_i V_\mu=\partial_i V_\mu-\omega^\nu_{\phantom{\nu}\mu i}V_\nu.
\]
For tensors with Lorentz indices, each index produces a term with the spin connection.
The spin connection transforms like a covariant vector under general coordinate transformations.
Consequently, the covariant derivative of a tetrad is zero:
\[
    \nabla_k e^i_\mu=\partial_k e^i_\mu+\Gamma^{i}_{jk}e^j_\mu-\omega^\nu_{\phantom{\nu}\mu k}e^i_\nu=0.
\]
Therefore, the covariant differentiation commutes with converting between coordinate and Lorentz indices.
This relation determines the spin connection in terms of the affine connection, the tetrad, and its partial derivatives.
The metricity of the affine connection leads to $\nabla_j g_{ik}=e^\mu_i e^\nu_k\nabla_j\eta_{\mu\nu}=-e^\mu_i e^\nu_k(\omega^\rho_{\phantom{\rho}\mu j}\eta_{\rho\nu}+\omega^\rho_{\phantom{\rho}\nu j}\eta_{\mu\rho})=-(\omega_{kij}+\omega_{ikj})=0$.
Consequently, $\nabla_i\eta_{\mu\nu}=0$ and the spin connection is antisymmetric in its first two indices: $\omega_{\mu\nu i}=-\omega_{\nu\mu i}$.
The spin connection $\mathring{\omega}_{ijk}$ constructed from $\mathring{\Gamma}^{j}_{ik}$ in the same way as $\omega_{ijk}$ constructed from $\Gamma^{j}_{ik}$ satisfies
\begin{equation}
    \omega_{ijk}=\mathring{\omega}_{ijk}+C_{ijk}
    \label{Riccirotation}
\end{equation}
and coincides with the Ricci rotation coefficients \cite{LL2}.\\

{\bf 3. Einstein--Cartan gravity}\\ \\
\noindent
The Einstein--Cartan theory, formulated by Sciama and Kibble \cite{EC,Lord,FranciscoNiko} and also known as the Einstein--Cartan--Sciama--Kibble theory, is the simplest theory of gravity that extends general relativity by relaxing the symmetry condition of the affine connection.
In this theory, the Lagrangian density for the gravitational field is given by $\mathfrak{L}_g=(-1/2\kappa)\mathfrak{e}R$, where $R=R_{ik}g^{ik}$ is the Ricci scalar,
\[
    R_{ik}=R^j_{\phantom{j}ijk}
\]
is the Ricci tensor, $\mathfrak{e}=\sqrt{-\mathfrak{g}}$ is the determinant of $e_i^\mu$ ($\mathfrak{g}$ is the determinant of $g_{ik}$), and $\kappa=8\pi G/c^4$.
In general relativity, the curvature tensor $R^l_{\phantom{l}ijk}$ reduces to the Riemann tensor $\mathring{R}^l_{\phantom{l}ijk}$.

The field equations of the Einstein--Cartan theory are obtained from the principle of least action with the variations of the metric and torsion tensors.
Equivalently, the field equations can be derived from variations of the tetrad and spin connection.
Varying the Lagrangian density of matter $\mathfrak{L}_m$ with respect to the spin connection gives the spin density:
\begin{equation}
    \mathfrak{S}_{\mu\nu}^{\phantom{\mu\nu}i}=-\mathfrak{S}_{\nu\mu}^{\phantom{\nu\mu}i}=2\delta\mathfrak{L}_m/\delta\omega^{\mu\nu}_{\phantom{\mu\nu}i}=\mathfrak{e}s_{\mu\nu}^{\phantom{\mu\nu}i},
    \label{spintensor}
\end{equation}
where $s_{\mu\nu}^{\phantom{\mu\nu}i}$ is the spin tensor.
The spin tensor is also determined by $\mathfrak{e}s_{ik}^{\phantom{ik}j}=2\delta\mathfrak{L}_m/\delta C^{ik}_{\phantom{ik}j}$ because of (\ref{Riccirotation}).
Varying the Lagrangian density of matter with respect to the tetrad gives the canonical energy--momentum density:
\[
    \mathfrak{T}^{\phantom{i}\mu}_i=\delta\mathfrak{L}_m/\delta e^i_\mu.
\]
The conservation law for spin density is $\nabla_k\mathfrak{S}_{ij}^{\phantom{ij}k}-2S_k\mathfrak{S}_{ij}^{\phantom{ij}k}=\mathfrak{T}_{ij}-\mathfrak{T}_{ji}$ \cite{Lord,FranciscoNiko}, where $S_k=S^i_{\phantom{i}ki}$ is the torsion vector.
Without spin, the energy--momentum tensor is symmetric, as in general relativity.
The conservation law for energy--momentum density is $\nabla_j\mathfrak{T}^{\phantom{i}j}_{i}-2S_j \mathfrak{T}^{\phantom{i}j}_i=2S^j_{\phantom{j}ki}\mathfrak{T}^{\phantom{j}k}_j+(1/2)R^{kl}_{\phantom{kl}ji}\mathfrak{S}_{kl}^{\phantom{kl}j}$.
Combining these two conservation laws gives the Belinfante--Rosenfeld equation \cite{Lord,FranciscoNiko}:
\begin{equation}
    T_{ik}=\frac{1}{\mathfrak{e}}\mathfrak{T}_{ik}-\frac{1}{2}(\nabla_j-2S_j)(s_{ik}^{\phantom{ik}j}-s_{k\phantom{j}i}^{\phantom{k}j}+s^j_{\phantom{j}ik}),
    \label{relation}
\end{equation}
where
\begin{equation}
    T_{ik}=2\delta\mathfrak{L}_m/\delta g^{ik}
    \label{energymomentummetric}
\end{equation}
is the metric energy--momentum tensor.

Equaling the variation of the total Lagrangian density for the gravitational field and matter with respect to the spin connection to zero gives the Cartan equations \cite{EC,Lord,FranciscoNiko}:
    \begin{equation}
    S^i_{\phantom{i}\mu\nu}-S_\mu e^i_\nu+S_\nu e^i_\mu=-\frac{\kappa}{2\mathfrak{e}}\mathfrak{S}^{\phantom{\mu\nu}i}_{\mu\nu},\quad S^j_{\phantom{j}ik}-S_i \delta^j_k+S_k \delta^j_i=-\frac{1}{2}\kappa s^{\phantom{ik}j}_{ik}.
    \label{Cartan}
\end{equation}
Equaling the variation of the total Lagrangian density for the gravitational field and matter with respect to the tetrad to zero gives the Einstein equations:
\[
    R^\mu_{\phantom{\mu}i}-\frac{1}{2}Re^\mu_i=\frac{\kappa}{\mathfrak{e}}\mathfrak{T}^{\phantom{i}\mu}_i.
\]
The Bianchi identity $\nabla_{[l}R^i_{\phantom{i}|n|jk]}=2R^i_{\phantom{i}nm[j}S^m_{\phantom{m}kl]}$ and the cyclic identity $R^m_{\phantom{m}[jkl]}=-2\nabla_{[l}S^m_{\phantom{m}jk]}+4S^m_{\phantom{m}n[j}S^n_{\phantom{n}kl]}$, where the square brackets denote antisymmetrization of indices except those between the bars, together with the Einstein and Cartan field equations give the conservation laws for the spin density and the energy--momentum density \cite{Lord,FranciscoNiko}

The Einstein equations are equivalent, because of (\ref{relation}), to
\begin{equation}
    G_{ik}=\mathring{R}_{ik}-\frac{1}{2}\mathring{R}g_{ik}=\kappa(T_{ik}+U_{ik}),
    \label{Einstein}
\end{equation}
where $\mathring{R}_{ik}$ and $\mathring{R}$ are constructed from $\mathring{\Gamma}^{j}_{ik}$ in the same way as $R_{ik}$ and $R$ are constructed from $\Gamma^{j}_{ik}$, $G_{ik}$ is the Einstein tensor, and
\begin{eqnarray}
    & & U_{ik}=\frac{1}{\kappa}\left(C^j_{\phantom{j}ij}C^l_{\phantom{l}kl}-C^l_{\phantom{l}ij}C^j_{\phantom{j}kl}-\frac{1}{2}g_{ik}(C^{jm}_{\phantom{jm}j}C^l_{\phantom{l}ml}-C^{mjl}C_{ljm})\right) \nonumber \\
    & & =\frac{1}{\kappa}\left((4S_i S_k-2S^j_{\phantom{j}il}S_{jk}^{\phantom{jk}l}-2S^j_{\phantom{j}il}S^l_{\phantom{l}kj}+S_{ijl}S_k^{\phantom{k}jl})-\frac{1}{2}g_{ik}(4S^j S_j-2S^l_{\phantom{l}mn}S^{nm}_{\phantom{nm}l}-S^l_{\phantom{l}mn}S_l^{\phantom{l}mn})\right). \nonumber
\end{eqnarray}
The Cartan equations (\ref{Cartan}) are equivalent to
\[
    S^k_{\phantom{k}ij}=-\frac{1}{2}\kappa(s_{ij}^{\phantom{ij}k}+\delta^k_{[i}s_{j]l}^{\phantom{j]l}l}),\quad C^k_{\phantom{k}ij}=\frac{1}{2}\kappa(s^k_{\phantom{k}ij}-s_{ij}^{\phantom{ij}k}-s_{j\phantom{k}i}^{\phantom{j}k}-g_{ij}s^{kl}_{\phantom{kl}l}+\delta^k_j s_{il}^{\phantom{il}l}),
\]
so putting torsion (\ref{torsion}) or contortion (\ref{contortion}) into the tensor $U_{ik}$ gives
\begin{equation}
    U^{ik}=\kappa\left(-s^{ij}_{\phantom{ij}[l}s^{kl}_{\phantom{kl}j]}-\frac{1}{2}s^{ijl}s^k_{\phantom{k}jl}+\frac{1}{4}s^{jli}s_{jl}^{\phantom{jl}k}+\frac{1}{8}g^{ik}(-4s^l_{\phantom{l}j[m}s^{jm}_{\phantom{jm}l]}+s^{jlm}s_{jlm})\right).
    \label{combined}
\end{equation}
This tensor represents the correction to the metric energy--momentum tensor from the spin contributions to the geometry of spacetime \cite{EC,Lord,FranciscoNiko}.
It is quadratic in the spin tensor, representing a spin-spin contact interaction.
Therefore, changing the signs of all components of the spin tensor does not affect this correction.
The spin tensor can also appear in $T_{ik}$ because $\mathfrak{L}_\textrm{m}$ can depend on torsion.

The Einstein--Cartan field equations relate the curvature of spacetime to the energy and momentum of matter and the torsion of spacetime to the spin angular momentum of matter.
The field equations (\ref{Einstein}) of the Einstein--Cartan theory of gravity, in which $U_{ik}$ is given by (\ref{combined}), have the same form as the standard Einstein equations of general relativity with the combined energy--momentum tensor $T_{ik}+U_{ik}$.
This theory is therefore equivalent to general relativity with spin, which contributes to the matter part of the equations.
This equivalence is possible because torsion is proportional to spin density and all terms containing torsion can be moved from the curvature side to the matter side of the field equations.

Without spin, the torsion tensor is equal to zero and the affine connection reduces to the Levi-Civita connection, given by the Christoffel symbols.
In this case, the Einstein--Cartan theory reduces to the general theory of relativity.
This theory is also indistinguishable in predictions from general relativity at densities of matter that are much lower than the Cartan density $\rho_\textrm{C}=m^2 c^4/G\hbar^2$ \cite{non}, which is about $10^{51}$kg/m$^3$ for the electron, so it passes all observational and experimental tests of general relativity \cite{EC}.\footnote{
Extending general relativity to the Einstein--Cartan theory may solve several problems in quantum field theory and cosmology.
Torsion imposes a spatial extension on fermions \cite{non} and removes the ultraviolet divergence of radiative corrections represented by loop Feynman diagrams \cite{torsional}.
Torsion also generates a negative correction from the spin-torsion coupling to the energy density, which acts like gravitational repulsion, which may prevent the formation of singularities in black holes and at the beginning of the Universe \cite{reg}.
Consequently, the collapsing matter in a black hole would avoid a singularity and instead reach a nonsingular bounce, after which it would expand as a new, closed universe on the other side of its event horizon \cite{universe}.
This avoidance is also valid for nonspherical collapse \cite{shear}.
Enormous particle pair production in strong nonstationary gravitational fields \cite{prod} can naturally generate a finite period of exponential expansion (inflation) \cite{infl}.
Accordingly, our Universe might have originated as a baby universe in a parent black hole existing in another universe \cite{Pat}.}\\

{\bf 4. Spinors and the Dirac equation}\\ \\
\noindent
The Dirac equation is a relativistic wave equation, which describes all massive spin-1/2 particles such as the electron and quarks \cite{Dirac,RQM,QFT}.
This equation is consistent with the principles of quantum mechanics and special theory of relativity.
It predicts the existence of the spin angular momentum of particles and the existence of antiparticles, and accounts for the fine structure of the hydrogen spectrum.

In flat spacetime, the Dirac equation for a free particle with mass $m$ is given by
\begin{equation}
    i\hbar\gamma^\mu\partial_\mu\psi=mc\psi,
    \label{Diracflat}
\end{equation}
where $\psi$ is the four-component wave function and $\gamma^\mu$ are the $4\times4$ Dirac matrices, satisfying the anticommutation relation:
\begin{equation}
    \{\gamma^\mu,\gamma^\nu\}=\gamma^\mu\gamma^\nu+\gamma^\nu\gamma^\mu=2\eta^{\mu\nu}I_4,
    \label{anticommutation}
\end{equation}
where $\eta^{\mu\nu}=\mbox{diag}(1,-1,-1,-1)$ is the Minkowski metric tensor and $I_4$ is the four-dimensional unit matrix.
Four is the lowest dimension for which the $\gamma^\mu$ can be defined.
In the presence of the electromagnetic field, $\partial_\mu$ is extended to $\partial_\mu+ieA_\mu$, where $A_\mu$ is the electromagnetic potential and $e$ is the electric charge of the particle.

In curved spacetime, the partial derivatives $\partial_\mu$ must be replaced with the covariant derivatives.
These derivatives can be determined in the spinor representation of the Lorentz group.
Let $L$ be a 4$\times$4 matrix such that
\begin{equation}
    \gamma^\mu=\Lambda^\mu_\nu L\gamma^\nu L^{-1}.
    \label{matrixL}
\end{equation}
This condition represents the constancy of the Dirac matrices under a Lorentz tetrad rotation combined with a similarity transformation and gives the matrix $L$ as a function of the Lorentz matrix $\Lambda^\mu_\nu$ \cite{Lord,FranciscoNiko,RQM,QFT}.
For an infinitesimal Lorentz transformation $\Lambda^\mu_\nu=\delta^\mu_\nu+\epsilon^\mu_{\phantom{\mu}\nu}$, where $\epsilon_{\mu\nu}=-\epsilon_{\nu\mu}$ are infinitesimal quantities, the solution for $L$ is
\[
    L=I_4+\frac{1}{2}\epsilon_{\mu\nu}\mathcal{G}^{\mu\nu},\quad L^{-1}=I_4-\frac{1}{2}\epsilon_{\mu\nu}\mathcal{G}^{\mu\nu},
\]
where $\mathcal{G}^{\mu\nu}$ are the generators of the spinor representation of the Lorentz group, proportional to the commutators of the Dirac matrices \cite{Lord,FranciscoNiko,RQM,QFT}:
\[
    \mathcal{G}^{\mu\nu}=\frac{1}{4}[\gamma^\mu,\gamma^\nu]=\frac{1}{4}(\gamma^\mu\gamma^\nu-\gamma^\nu\gamma^\mu).
\]
The matrices $L$ compose the Lorentz group in spinor representation. 

A spinor $\psi$ and its adjoint $\overline{\psi}$ are defined as quantities that transform according to \cite{Lord,FranciscoNiko}
\begin{equation}
    \tilde{\psi}=L\psi,\quad \overline{\tilde{\psi}}=\overline{\psi}L^{-1}.
    \label{spinorL}
\end{equation}
Accordingly, the product $\overline{\psi}\psi$ is a scalar: $\overline{\tilde{\psi}}\tilde{\psi}=\overline{\psi}\psi$.
The transformation law of the Dirac matrices shows that they can be regarded as quantities that have, in addition to the Lorentz vector index $\mu$, one spinor index and one adjoint-spinor index.
The product $\psi\overline{\psi}$ transforms like the Dirac matrices: $\tilde{\psi}\overline{\tilde{\psi}}=L\psi\overline{\psi}L^{-1}$.
The spinors $\psi$ and $\overline{\psi}$ can be used to construct bilinear forms that are linear both in $\psi$ and $\overline{\psi}$ and transform like tensors.
For example, $\overline{\psi}\gamma^\mu\psi$ transforms like a contravariant Lorentz vector: $\overline{\psi}\gamma^\mu\psi\rightarrow\overline{\psi}L^{-1}\Lambda^\mu_\nu L\gamma^\nu L^{-1}L\psi=\Lambda^\mu_\nu \overline{\psi}\gamma^\nu\psi$.

For an infinitesimal Lorentz transformation, the Hermitian conjugate of $L$ is $L^\dag=I_4+(1/8)\epsilon_{\mu\nu}(\gamma^{\nu\dag}\gamma^{\mu\dag}-\gamma^{\mu\dag}\gamma^{\nu\dag})$.
The relations $\gamma^{0\dag}=\gamma^0$ and $\gamma^{\alpha\dag}=-\gamma^\alpha$, where $\alpha\in\{1,2,3\}$, together with the definition of $\gamma^\mu$ give $L^\dag\gamma^0=\gamma^0 L^{-1}$.
Therefore, the quantity $\psi^\dag\gamma^0$ transforms like an adjoint spinor: $\psi^\dag\gamma^0\rightarrow\psi^\dag L^\dag\gamma^0=\psi^\dag\gamma^0 L^{-1}$.
Accordingly, the adjoint spinor is related to the conjugate of the spinor:
\begin{equation}
    \overline{\psi}=\psi^\dag\gamma^0.
    \label{adjointL}
\end{equation}

The derivative of a spinor does not transform like a spinor: $\partial_i\tilde{\psi}=L\partial_i\psi+\partial_i L\psi$.
The spinor connection $\Gamma_i$ that transforms according to
\[
    \tilde{\Gamma}_i=L\Gamma_i L^{-1}+\partial_i LL^{-1}
\]
is used to construct the covariant derivative of a spinor \cite{Lord,FranciscoNiko}:
\begin{equation}
    \nabla_i\psi=\partial_i\psi-\Gamma_i \psi,
    \label{covariant1}
\end{equation}
which transforms like a spinor: $\nabla_i\tilde{\psi}=\partial_i\tilde{\psi}-\tilde{\Gamma}_i\tilde{\psi}=L\partial_i\psi+\partial_i L\psi-(L\Gamma_i L^{-1}+\partial_i LL^{-1})L\psi=L\nabla_i\psi$.
Because $\overline{\psi}\psi$ is a scalar, $\nabla_i(\overline{\psi}\psi)=\partial_i(\overline{\psi}\psi)$, the chain rule for covariant differentiation gives the covariant derivative of an adjoint spinor:
\begin{equation}
    \nabla_i\overline{\psi}=\partial_i\overline{\psi}+\overline{\psi}\Gamma_i.
    \label{covariant2}
\end{equation}

The Dirac matrices $\gamma^\mu$ transform like $\psi\overline{\psi}$, whose covariant derivative is $\nabla_i(\psi\overline{\psi})=
\nabla_i\psi\overline{\psi}+\psi\nabla_i\overline{\psi}=\partial_i(\psi\overline{\psi})-\Gamma_i \psi\overline{\psi}+\psi\overline{\psi}\Gamma_i=\partial_i(\psi\overline{\psi})-[\Gamma_i,\psi\overline{\psi}]$.
Therefore, the covariant derivative of a Dirac matrix is \cite{Lord,FranciscoNiko}
\[
    \nabla_i\gamma^\mu=\omega^{\mu}_{\phantom{\mu}\nu i}\gamma^\nu-[\Gamma_i,\gamma^\mu].
\]
A quantity $\overline{\psi}\gamma^i\nabla_i\psi$, where $\gamma^i=e^i_\mu\gamma^\mu$, transforms under Lorentz rotations (and general coordinate transformations) like a scalar:
\[
    \overline{\psi}\gamma^i\nabla_i\psi\rightarrow\overline{\psi}L^{-1}L\gamma^i L^{-1}L\nabla_i\psi=\overline{\psi}\gamma^i\nabla_i\psi.
\]
So does a quantity $\nabla_i\overline{\psi}\gamma^i\psi$.
The difference $\overline{\psi}\gamma^i\nabla_i\psi-\nabla_i\overline{\psi}\gamma^i\psi$ is imaginary.
It can be used to construct the simplest Lagrangian density for a Dirac field that contains its covariant derivatives.

The relation $\nabla_i\eta_{\mu\nu}=0$ gives $\nabla_i\gamma^\mu=0$ because the Dirac matrices $\gamma^\mu$ depend only on the tensor $\eta_{\mu\nu}$.
Consequently, $\gamma_\mu\nabla_i\gamma^\mu=\omega_{\mu\nu i}\gamma^\mu \gamma^\nu-\gamma_\mu \Gamma_i \gamma^\mu+4\Gamma_i=0$.
Its solution is $\Gamma_i=-(1/4)\omega_{\mu\nu i}\gamma^\mu \gamma^\nu-A_i I_4$, where $A_i$ is a covariant vector, which could be proportional to the electromagnetic potential.
Therefore, the spinor connection $\Gamma_i$ is given, up to the addition of an arbitrary vector multiple of the unit matrix, by the Fock--Ivanenko coefficients \cite{FockIvanenko,Lord,FranciscoNiko}:
\begin{equation}
    \Gamma_i=-\frac{1}{4}\omega_{\mu\nu i}\gamma^\mu \gamma^\nu=-\frac{1}{2}\omega_{\mu\nu i}\mathcal{G}^{\mu\nu}.
    \label{FI}
\end{equation}
The relation (\ref{Riccirotation}) gives
\begin{equation}
    \nabla_i\psi=\mathring{\nabla}_i\psi+\frac{1}{4}C_{\mu\nu i}\gamma^\mu \gamma^\nu\psi,\quad \nabla_i\overline{\psi}=\mathring{\nabla}_i\overline{\psi}-\frac{1}{4}C_{\mu\nu i}\overline{\psi}\gamma^\mu \gamma^\nu.
    \label{derivativecontortion}
\end{equation}
The curvature tensor with two Lorentz and two coordinate indices depends on the spin connection and its partial derivatives: $R^\mu_{\phantom{\mu}\nu ik}=\omega^\mu_{\phantom{\mu}\nu k,i}-\omega^\mu_{\phantom{\mu}\nu i,k}+\omega^\rho_{\phantom{\rho}\nu k}\omega^\mu_{\phantom{\mu}\rho i}-\omega^\rho_{\phantom{\rho}\nu i}\omega^\mu_{\phantom{\mu}\rho k}$.
The commutator of the covariant derivatives of a spinor is $\nabla_i\nabla_j\psi-\nabla_j\nabla_i\psi=K_{ij}\psi+2S^k_{\phantom{k}ij}\nabla_k\psi$, where $K_{ij}=\partial_j\Gamma_i-\partial_i\Gamma_j+[\Gamma_i,\Gamma_j]=(1/4)R_{\mu\nu ij}\gamma^\mu \gamma^\nu=(1/2)R_{\mu\nu ij}\mathcal{G}^{\mu\nu}$ is the curvature spinor, which transforms like a quantity with one spinor index and one adjoint-spinor index: $\tilde{K}_{ij}=LK_{ij}L^{-1}$ \cite{Lord,FranciscoNiko}.\\

{\bf 5. The spin tensor and the Dirac equation with torsion}\\ \\
The Dirac Lagrangian density for a spinor field, representing a particle in a gravitational field, is given by
\begin{equation}
    \mathfrak{L}_\psi=\frac{1}{2}i\hbar c\mathfrak{e}e^i_\mu(\overline{\psi}\gamma^\mu\nabla_i\psi-\nabla_i\overline{\psi}\gamma^\mu\psi)-mc^2\mathfrak{e}\overline{\psi}\psi.
    \label{Lagrangian}
\end{equation}
Using the covariant derivatives (\ref{covariant1}) and (\ref{covariant2}), this Lagrangian is equal to
\[
    \mathfrak{L}_\psi=\frac{1}{2}i\hbar c\mathfrak{e}e^i_\mu(\overline{\psi}\gamma^\mu\partial_i\psi-\partial_i\overline{\psi}\gamma^\mu\psi)-\frac{1}{2}i\hbar c\mathfrak{e}e^i_\mu\overline{\psi}\{\gamma^\mu,\Gamma_i\}\psi-mc^2\mathfrak{e}\overline{\psi}\psi.
\]
The term with the spinor connection is
\[
    \{\gamma^\mu,\Gamma_i\}=-\frac{1}{4}\omega_{\nu\rho i}\{\gamma^\mu,\gamma^\nu \gamma^\rho\}=-\frac{1}{4}\omega_{\nu\rho i}\{\gamma^\mu,\gamma^{[\nu}\gamma^{\rho]}\}=-\frac{1}{2}\omega_{\nu\rho i}\gamma^{[\mu} \gamma^\nu \gamma^{\rho]},
\]
so (\ref{Lagrangian}) becomes
\begin{equation}
    \mathfrak{L}_\psi=\frac{1}{2}i\hbar c\mathfrak{e}e^i_\mu(\overline{\psi}\gamma^\mu\partial_i\psi-\partial_i\overline{\psi}\gamma^\mu\psi)+\frac{1}{4}i\hbar c\mathfrak{e}\omega_{\nu\rho\mu}\overline{\psi}\gamma^{[\mu} \gamma^\nu \gamma^{\rho]}\psi-mc^2\mathfrak{e}\overline{\psi}\psi.
    \label{Lagrangianspin}
\end{equation}

The spin density (\ref{spintensor}) corresponding to the Dirac Lagrangian (\ref{Lagrangianspin}) is \cite{HehlDatta,Lord,FranciscoNiko}
\[
    \mathfrak{S}^{\mu\nu\rho}=\frac{1}{2}i\hbar c\mathfrak{e}\overline{\psi}\gamma^{[\mu}\gamma^\nu\gamma^{\rho]}\psi=\mathfrak{S}^{[\mu\nu\rho]},
\]
which is completely antisymmetric and does not depend on the spinor mass $m$.
The corresponding spin tensor is
\begin{equation}
    s^{\mu\nu\rho}=\frac{1}{2}i\hbar c\overline{\psi}\gamma^{[\mu}\gamma^\nu\gamma^{\rho]}\psi=s^{[\mu\nu\rho]}=-\frac{1}{2}\hbar ce^{\mu\nu\rho\sigma}j^5_\sigma,
    \label{spintensorDirac}
\end{equation}
where
\begin{equation}
    j^{5\mu}=\overline{\psi}\gamma^\mu\gamma^5\psi
    \label{axialcurrent}
\end{equation}
is the axial current density pseudovector, $e^{\mu\nu\rho\sigma}$ is the completely antisymmetric unit pseudotensor, and $\gamma^5=i\gamma^0\gamma^1\gamma^2\gamma^3$ is the chirality matrix.
The Cartan equations (\ref{Cartan}) for the spin tensor (\ref{spintensorDirac}) give a completely antisymmetric torsion tensor:
\begin{equation}
    S_{\mu\nu\rho}=C_{\mu\nu\rho}=-\frac{1}{4}i\kappa\hbar c\overline{\psi}\gamma_{[\mu}\gamma_\nu\gamma_{\rho]}\psi=\frac{1}{4}{\kappa}\hbar ce_{\mu\nu\rho\sigma}j^{5\sigma},\quad S_i=0.
    \label{torsionDirac}
\end{equation}

Varying the Lagrangian (\ref{Lagrangian}) with respect to $\overline{\psi}$ and $\psi$ gives the Dirac equation and its adjoint \cite{Lord,FranciscoNiko,HehlDatta}:
\begin{equation}
    i\hbar \gamma^\mu e_\mu^i \nabla_i \psi = mc\psi,\quad -i\hbar \nabla_i\overline{\psi} \gamma^\mu e_\mu^i= mc\overline{\psi}.
    \label{Dirac}
\end{equation}
These equations generalize the Dirac equation (\ref{Diracflat}) to a curved spacetime, in which the covariant derivatives (\ref{covariant1}) and (\ref{covariant2}) are used.
In terms of the Fock--Ivanenko coefficients, this equation is
\[
    i\hbar \gamma^\mu e_\mu^i (\partial_i \psi + \frac{1}{4}\omega_{\rho\sigma i}\gamma^\rho \gamma^\sigma) = mc\psi.
\]
Putting (\ref{torsionDirac}) into (\ref{derivativecontortion}) yields
\begin{eqnarray}
& & 
    \gamma^\mu e^k_\mu\nabla_k\psi=\gamma^\mu e^k_\mu\mathring{\nabla}_k\psi+\frac{1}{16}\kappa\hbar ce_{\mu\nu\rho\tau}(\overline{\psi}\gamma^\tau\gamma^5\psi)\gamma^\rho\gamma^\mu\gamma^\nu\psi=\gamma^\mu e^k_\mu\mathring{\nabla}_k\psi+\frac{1}{16}i\kappa\hbar ce_{\mu\nu\rho\tau}(\overline{\psi}\gamma^\tau\gamma^5\psi)e^{\mu\nu\rho\sigma}\gamma_\sigma\gamma^5\psi \nonumber \\
    & & =\gamma^\mu e^k_\mu\mathring{\nabla}_k\psi-\frac{3}{8}i\kappa\hbar c(\overline{\psi}\gamma^\mu\gamma^5\psi)\gamma_\mu\gamma^5\psi. \nonumber
\end{eqnarray}
Therefore, the Dirac equation (\ref{Dirac}) becomes the Hehl--Datta equation \cite{Lord,FranciscoNiko,HehlDatta}:
\begin{equation}
    i\hbar\gamma^\mu e^k_\mu\mathring{\nabla}_k\psi+\mathcal{K}(\overline{\psi}\gamma^\mu\gamma^5\psi)\gamma_\mu\gamma^5\psi=mc\psi,\quad -i\hbar e^k_\mu\mathring{\nabla}_k\overline{\psi}\gamma^\mu+\mathcal{K}(\overline{\psi}\gamma^\mu\gamma^5\psi)\overline{\psi}\gamma_\mu\gamma^5=mc\overline{\psi},
    \label{Diracfinal}
\end{equation}
where
\[
\mathcal{K}=\frac{3}{8}\kappa\hbar^2 c.
\]
These equations are cubic in spinor fields.
The Dirac equation (\ref{Diracfinal}) can be written in a shorter form as
\begin{equation}
    i\hbar D\psi=mc\psi-\mathcal{K}j^{5\mu}\gamma_\mu\gamma^5\psi, \quad D\psi=\gamma^\mu e^i_\mu\mathring{\nabla}_i\psi,
    \label{DiracD}
\end{equation}
where $D$ is the Dirac operator.\\

{\bf 6. The energy--momentum tensor for a spinor field}\\ \\
Varying the Lagrangian density (\ref{Lagrangian}) with respect to $e^i_a$ gives the canonical energy--momentum density for a spinor field \cite{Lord,FranciscoNiko}:
\begin{equation}
    \mathfrak{T}^{\phantom{i}\mu}_i=\frac{1}{2}i\hbar c\mathfrak{e}(\overline{\psi}\gamma^\mu\nabla_i\psi-\nabla_i\overline{\psi}\gamma^\mu\psi-e^\mu_i\overline{\psi}\gamma^j\nabla_j\psi+e^\mu_i\nabla_j\overline{\psi}\gamma^j\psi)+mc^2\mathfrak{e}e^\mu_i\overline{\psi}\psi.
    \label{energymomentumDirac}
\end{equation}
The metric energy--momentum tensor for this field can be obtained by putting (\ref{energymomentumDirac}) in the Belinfante--Rosenfeld equation (\ref{relation}) with the spin tensor (\ref{spintensorDirac}) and the torsion tensor (\ref{torsionDirac}), or by varying (\ref{Lagrangian}) with respect to $g^{ik}$ according to (\ref{energymomentummetric}).
Using the identity
\[
    \delta\gamma^j/\delta g^{ik}=\frac{1}{2}\delta^j_{(i}\gamma_{k)},
\]
which follows from $\gamma^i\gamma^k+\gamma^k\gamma^i=2g^{ik}I_4$, equivalent to (\ref{anticommutation}), where $\gamma_i=e^\mu_i\gamma_\mu$, leads to
\[
    T_{ik}=\frac{1}{2}i\hbar c(\overline{\psi}\gamma_{(i}\nabla_{k)}\psi-\nabla_{(i}\overline{\psi}\gamma_{k)}\psi-g_{ik}\overline{\psi}\gamma^j\nabla_j\psi+g_{ik}\nabla_j\overline{\psi}\gamma^j\psi)+mc^2 g_{ik}\overline{\psi}\psi.
\]
This equation reduces, using the Dirac equation (\ref{Dirac}), to \cite{FranciscoNiko,Kerlick}
\[
    T_{ik}=\frac{1}{2}i\hbar c(\overline{\psi}\delta^j_{(i}\gamma_{k)}\nabla_j\psi-\nabla_j\overline{\psi}\delta^j_{(i}\gamma_{k)}\psi),
\]
which can be written in terms of the Levi-Civita connection, using (\ref{derivativecontortion}) and (\ref{torsionDirac}):
\[
    T_{ik}=\frac{1}{2}i\hbar c(\overline{\psi}\delta^j_{(i}\gamma_{k)}\mathring{\nabla}_j\psi-\mathring{\nabla}_j\overline{\psi}\delta^j_{(i}\gamma_{k)}\psi)+\frac{1}{8}\kappa(\hbar c)^2(-j^5_i j^5_k+j^{5l} j^5_l g_{ik}).
\]
Putting the completely antisymmetric spin tensor (\ref{spintensorDirac}) in (\ref{combined}) gives
\[
    U^{ik}=\frac{1}{4}\kappa\left(s^{ijl}s^k_{\phantom{k}jl}-\frac{1}{2}g^{ik}s^{jlm}s_{jlm}\right)=\frac{1}{16}\kappa(\hbar c)^2(2j^{5i} j^{5k}+j^{5l} j^5_l g^{ik}).
\]
The combined energy--momentum tensor for a Dirac field is thus \cite{FranciscoNiko,Kerlick}
\begin{eqnarray}
    & & T_{ik}+U_{ik}=\frac{1}{2}i\hbar c(\overline{\psi}\delta^j_{(i}\gamma_{k)}\mathring{\nabla}_j\psi-\mathring{\nabla}_j\overline{\psi}\delta^j_{(i}\gamma_{k)}\psi)+\frac{3}{16}\kappa(\hbar c)^2 j^{5l} j^5_l g_{ik} \nonumber \\
    & & =\frac{1}{4}i\hbar c\left[\overline{\psi}\gamma_\mu(e^\mu_k\mathring{\nabla}_i\psi+e^\mu_i\mathring{\nabla}_k\psi)-(e^\mu_k\mathring{\nabla}_i\overline{\psi}+e^\mu_i\mathring{\nabla}_k\overline{\psi})\gamma_\mu\psi\right]+\frac{1}{2}\mathcal{K}c j^{5\mu} j^5_\mu g_{ik}.
    \label{combinedDirac}
\end{eqnarray}\\

{\bf 7. The Dirac equation in a centrally symmetric gravitational field}\\ \\
The centrally symmetric gravitational field is given by the metric \cite{LL2}
\begin{equation}
    ds^2=e^{\nu(t,r)}c^2 dt^2-e^{\lambda(t,r)}dr^2-r^2 d\theta^2-r^2\sin^2\theta\,d\phi^2,
    \label{centralmetric}
\end{equation}
where $\nu$ and $\lambda$ are functions of time $t$ and radial coordinate $r$.
The nonzero components of the metric tensor and the inverse metric tensor are $
g_{tt}=e^\nu, g_{rr}=-e^\lambda, g_{\theta\theta}=-r^2, g_{\phi\phi}=-r^2\sin^2\theta, g^{tt}=e^{-\nu}, g^{rr}=-e^{-\lambda}, g^{\theta\theta}=-1/r^2, g^{\phi\phi}=-1/r^2\sin^2\theta$.
The nonzero Christoffel symbols (\ref{affineconnection}) for this metric are \cite{LL2}
\begin{eqnarray}
    & & \mathring{\Gamma}^t_{tt}=\frac{1}{2}\dot{\nu},\quad \mathring{\Gamma}^t_{tr}=\frac{1}{2}\nu',\quad \mathring{\Gamma}^t_{rr}=\frac{1}{2}\dot{\lambda}e^{\lambda-\nu},\quad \mathring{\Gamma}^r_{tt}=\frac{1}{2}\nu'e^{\nu-\lambda},\quad \mathring{\Gamma}^r_{tr}=\frac{1}{2}\dot{\lambda},\quad \quad \mathring{\Gamma}^r_{rr}=\frac{1}{2}\lambda', \nonumber \\
    & & \mathring{\Gamma}^r_{\theta\theta}=-re^{-\lambda},\quad \mathring{\Gamma}^r_{\phi\phi}=-r\sin^2\theta\,e^{-\lambda},\quad \mathring{\Gamma}^\theta_{r\theta}=1/r,\quad \mathring{\Gamma}^\theta_{\phi\phi}=-\sin\theta\cos\theta,\quad \mathring{\Gamma}^\phi_{r\phi}=1/r,\quad \mathring{\Gamma}^\phi_{\theta\phi}=\cot\theta. \nonumber
\end{eqnarray}
The tetrad can be chosen as
\[
    e_0^t=e^{-\nu/2},\quad e_3^r=e^{-\lambda/2},\quad e_1^\theta=1/r,\quad e_2^\phi=1/r\sin\theta,
\]
and the corresponding inverse tetrad as
\[
    e^0_t=e^{\nu/2},\quad e^3_r=e^{\lambda/2},\quad e^1_\theta=r,\quad e^2_\phi=r\sin\theta.
\]
Consequently, the nonzero components of the spin connection (\ref{spinconnectiontetrad}) are \cite{rotations}
\begin{eqnarray}
    & & \omega^0_{\phantom{0}3t}=e^0_t(e^t_{3,t}+\mathring{\Gamma}^t_{rt}e^r_3)=e^{\nu/2}\left(\frac{1}{2}\nu'\right)e^{-\lambda/2}=\frac{1}{2}\nu'e^{(\nu-\lambda)/2}=\omega^3_{\phantom{3}0t}, \nonumber \\
    & & \omega^0_{\phantom{0}3r}=e^0_t(e^t_{3,r}+\mathring{\Gamma}^t_{rr}e^r_3)=e^{\nu/2}\left(\frac{1}{2}\dot{\lambda}e^{\lambda-\nu}\right)e^{-\lambda/2}=\frac{1}{2}\dot{\lambda}e^{(\lambda-\nu)/2}=\omega^3_{\phantom{3}0r}, \nonumber \\
    & & \omega^3_{\phantom{3}1\theta}=e^3_r(e^r_{1,\theta}+\mathring{\Gamma}^r_{\theta\theta}e^\theta_1)=e^{\lambda/2}(-re^{-\lambda})1/r=-e^{-\lambda/2}=-\omega^1_{\phantom{1}3\theta}, \nonumber \\
    & & \omega^3_{\phantom{3}2\phi}=e^3_r(e^r_{2,\phi}+\mathring{\Gamma}^r_{\phi\phi}e^\phi_2)=e^{\lambda/2}(-r\sin^2\theta\,e^{-\lambda})1/r\sin\theta=-\sin\theta\,e^{-\lambda/2}=-\omega^2_{\phantom{2}3\phi}, \nonumber \\
    & & \omega^1_{\phantom{1}2\phi}=e^1_\theta(e^\theta_{2,\phi}+\mathring{\Gamma}^\theta_{\phi\phi}e^\phi_2)=r(-\sin\theta\cos\theta)1/r\sin\theta=-\cos\theta=-\omega^2_{\phantom{2}1\phi}. \nonumber
    \label{spinconnection}
\end{eqnarray}

The nonzero components of the spinor connection (\ref{FI}) are \cite{rotations}
\begin{eqnarray}
    & & \Gamma_t=-\frac{1}{4}\omega_{\mu\nu t}\gamma^{\mu}\gamma^{\nu}=-\frac{1}{2}\omega_{03t}\gamma^0\gamma^3=-\frac{1}{4}\nu'e^{(\nu-\lambda)/2}\gamma^0\gamma^3, \nonumber \\
    & & \Gamma_r=-\frac{1}{4}\omega_{\mu\nu r}\gamma^{\mu}\gamma^{\nu}=-\frac{1}{2}\omega_{03r}\gamma^0\gamma^3=-\frac{1}{4}\dot{\lambda}e^{(\lambda-\nu)/2}\gamma^0\gamma^3, \nonumber \\
    & & \Gamma_\theta=-\frac{1}{4}\omega_{\mu\nu \theta}\gamma^{\mu}\gamma^{\nu}=-\frac{1}{2}\omega_{31\theta}\gamma^3\gamma^1=-\frac{1}{2}e^{-\lambda/2}\gamma^3\gamma^1, \nonumber \\
    & & \Gamma_\phi=-\frac{1}{4}\omega_{\mu\nu \phi}\gamma^{\mu}\gamma^{\nu}=-\frac{1}{2}\omega_{32\phi}\gamma^3\gamma^2-\frac{1}{2}\omega_{12\phi}\gamma^1\gamma^2=-\frac{1}{2}\sin\theta\,e^{-\lambda/2}\gamma^3\gamma^2-\frac{1}{2}\cos\theta\,\gamma^1\gamma^2.
    \label{spinorconnection}
\end{eqnarray}
Therefore, the covariant derivative terms in the Dirac equation are
\begin{eqnarray}
    & & \gamma^0 e^t_0\nabla_t\psi=\gamma^0 e^t_0(\partial_{ct}-\Gamma_t)\psi=\gamma^0 e^{-\nu/2}\left(\partial_{ct}\psi+\frac{1}{4}\nu'e^{(\nu-\lambda)/2}\gamma^0\gamma^3\psi\right), \nonumber \\
    & & \gamma^3 e^r_3\nabla_r\psi=\gamma^3 e^r_3(\partial_r-\Gamma_r)\psi=\gamma^3 e^{-\lambda/2}\left(\partial_r\psi+\frac{1}{4}\dot{\lambda}e^{(\lambda-\nu)/2}\gamma^0\gamma^3\psi\right), \nonumber \\
    & & \gamma^1 e^\theta_1\nabla_\theta\psi=\gamma^1 e^\theta_1(\partial_\theta-\Gamma_\theta)\psi=\gamma^1 \frac{1}{r}\left(\partial_\theta\psi+\frac{1}{2}e^{-\lambda/2}\gamma^3\gamma^1\psi\right), \nonumber \\
    & & \gamma^2 e^\phi_2\nabla_\phi\psi=\gamma^2 e^\phi_2(\partial_\phi-\Gamma_\phi)\psi=\gamma^2 \frac{1}{r\sin\theta}\left(\partial_\phi\psi+\frac{1}{2}\sin\theta\,e^{-\lambda/2}\gamma^3\gamma^2+\frac{1}{2}\cos\theta\,\gamma^1\gamma^2\psi\right), \nonumber
\end{eqnarray}
giving the Dirac operator acting on a spinor $\psi$:
\begin{eqnarray}
    & & D\psi=\gamma^\mu e^i_\mu\nabla_i\psi=\gamma^0 e^t_0\nabla_t\psi+\gamma^3 e^r_3\nabla_r\psi+\gamma^1 e^\theta_1\nabla_\theta\psi+\gamma^2 e^\phi_2\nabla_\phi\psi \nonumber \\
    & & =e^{-\nu/2}\gamma^0\partial_{ct}\psi+\frac{1}{4}\nu'e^{-\lambda/2}\gamma^3\psi+e^{-\lambda/2}\gamma^3\partial_r\psi+\frac{1}{4}\dot{\lambda}e^{-\nu/2}\gamma^0\psi \nonumber \\
    & & +\frac{1}{r}\gamma^1\partial_\theta\psi+\frac{1}{2r}e^{-\lambda/2}\gamma^3\psi
    +\frac{1}{r\sin\theta}\gamma^2\partial_\phi\psi+\frac{1}{2r}e^{-\lambda/2}\gamma^3\psi+\frac{1}{2r}\cot\theta\,\gamma^1\psi \nonumber \\
    & & =e^{-\nu/2}\gamma^0\left(\partial_{ct}+\frac{1}{4}\dot{\lambda}\right)\psi+e^{-\lambda/2}\gamma^3\left(\partial_r+\frac{1}{r}+\frac{1}{4}\nu'\right)\psi+\frac{1}{r}\gamma^1\left(\partial_\theta+\frac{1}{2}\cot\theta\right)\psi+\frac{1}{r\sin\theta}\gamma^2\partial_\phi\psi.
    \label{Diracoperator}
\end{eqnarray}
The Dirac equation with torsion (\ref{DiracD}) in a centrally symmetric gravitational field is given by
\begin{eqnarray}
    & & i\hbar\left[e^{-\nu/2}\gamma^0\left(\partial_{ct}+\frac{1}{4}\dot{\lambda}\right)\psi+e^{-\lambda/2}\gamma^3\left(\partial_r+\frac{1}{r}+\frac{1}{4}\nu'\right)\psi+\frac{1}{r}\gamma^1\left(\partial_\theta+\frac{1}{2}\cot\theta\right)\psi+\frac{1}{r\sin\theta}\gamma^2\partial_\phi\psi\right] \nonumber \\
    & & =mc\psi-\mathcal{K}j^{5\mu}\gamma_\mu\gamma^5\psi.
    \label{Diracspherical}
\end{eqnarray}\\

{\bf 8. Rotation of the Dirac matrices and the angular Dirac operator}\\ \\
A unitary operator \cite{rotations}
\begin{eqnarray}
    & & U=\exp\left(\frac{1}{2}\phi\gamma^1\gamma^2\right)\exp\left(\frac{1}{2}\theta\gamma^3\gamma^1\right)=\exp\left(-\frac{i}{2}\phi\Sigma_3\right)\exp\left(-\frac{i}{2}\theta\Sigma_2\right), \nonumber \\
    & & U^\dagger=U^{-1}=\exp\left(-\frac{1}{2}\theta\gamma^3\gamma^1\right)\exp\left(-\frac{1}{2}\phi\gamma^1\gamma^2\right)=\exp\left(\frac{i}{2}\theta\Sigma_2\right)\exp\left(\frac{i}{2}\phi\Sigma_3\right),
    \label{unitary}
\end{eqnarray}
where $\Sigma_\mu$ are the spin matrices, generates a rotation of four-component Dirac spinors \cite{RQM,QFT}.
The spinor and its adjoint rotate according to (\ref{spinorL}):
\[
    \tilde{\psi}=U\psi,\quad \overline{\tilde{\psi}}=\overline{\psi}U^{-1}.
\]
Under this rotation, the Dirac matrices transform according to\footnote{
Because a Dirac matrix commutes with the product of two other Dirac matrices, it commutes with the exponential function of that product.
Because a Dirac matrix anticommutes with the product of itself and another Dirac matrix, changing the order of multiplication of this matrix and the exponential function of that product changes the signs of odd powers of that product in the power series but keeps the signs of even powers, so the sign of the argument of the exponential function is changed.
}
\begin{eqnarray}
    & & \tilde{\gamma}^3=U\gamma^3U^{-1}=U\gamma^3\exp\left(-\frac{1}{2}\theta\gamma^3\gamma^1\right)\exp\left(-\frac{1}{2}\phi\gamma^1\gamma^2\right)=U\exp\left(\frac{1}{2}\theta\gamma^3\gamma^1\right)\exp\left(-\frac{1}{2}\phi\gamma^1\gamma^2\right)\gamma^3 \nonumber \\
    & & =\exp\left(\frac{1}{2}\phi\gamma^1\gamma^2\right)\exp\left(\theta\gamma^3\gamma^1\right)\exp\left(-\frac{1}{2}\phi\gamma^1\gamma^2\right)\gamma^3 \nonumber \\
    & & =\left(\cos\frac{\phi}{2}+\sin\frac{\phi}{2}\gamma^1\gamma^2\right)(\cos\theta+\sin\theta\,\gamma^3\gamma^1)\left(\cos\frac{\phi}{2}-\sin\frac{\phi}{2}\gamma^1\gamma^2\right)\gamma^3= \nonumber \\
    & & =\left(\cos\theta\cos^2\frac{\phi}{2}-\cos\theta\cos\frac{\phi}{2}\sin\frac{\phi}{2}\gamma^1\gamma^2+\sin\theta\cos^2\frac{\phi}{2}\gamma^3\gamma^1-\sin\theta\cos\frac{\phi}{2}\sin\frac{\phi}{2}\gamma^2\gamma^3\right)\gamma^3 \nonumber \\
    & & +\left(\cos\theta\sin\frac{\phi}{2}\cos\frac{\phi}{2}\gamma^1\gamma^2+\cos\theta\sin^2\frac{\phi}{2}-\sin\theta\sin\frac{\phi}{2}\cos\frac{\phi}{2}\gamma^2\gamma^3+\sin\theta\sin^2\frac{\phi}{2}\gamma^1\gamma^3\right)\gamma^3 \nonumber \\
    & & =(\cos\theta+\sin\theta\cos\phi\,\gamma^3\gamma^1-\sin\theta\sin\phi\,\gamma^2\gamma^3)\gamma^3=\sin\theta\cos\phi\,\gamma^1+\sin\theta\sin\phi\,\gamma^2+\cos\theta\,\gamma^3=\gamma^r={\bm\gamma}\cdot{\bf n}_r,
    \label{gamma3}
\end{eqnarray}
\begin{eqnarray}
    & & \tilde{\gamma}^1=U\gamma^1U^{-1}=U\gamma^1\exp\left(-\frac{1}{2}\theta\gamma^3\gamma^1\right)\exp\left(-\frac{1}{2}\phi\gamma^1\gamma^2\right)=U\exp\left(\frac{1}{2}\theta\gamma^3\gamma^1\right)\exp\left(\frac{1}{2}\phi\gamma^1\gamma^2\right)\gamma^1 \nonumber \\
    & & =\exp\left(\frac{1}{2}\phi\gamma^1\gamma^2\right)\exp\left(\theta\gamma^3\gamma^1\right)\exp\left(\frac{1}{2}\phi\gamma^1\gamma^2\right)\gamma^1 \nonumber \\
    & & =\left(\cos\frac{\phi}{2}+\sin\frac{\phi}{2}\gamma^1\gamma^2\right)(\cos\theta+\sin\theta\,\gamma^3\gamma^1)\left(\cos\frac{\phi}{2}+\sin\frac{\phi}{2}\gamma^1\gamma^2\right)\gamma^1= \nonumber \\
    & & =\left(\cos\theta\cos^2\frac{\phi}{2}+\cos\theta\cos\frac{\phi}{2}\sin\frac{\phi}{2}\gamma^1\gamma^2+\sin\theta\cos^2\frac{\phi}{2}\gamma^3\gamma^1+\sin\theta\cos\frac{\phi}{2}\sin\frac{\phi}{2}\gamma^2\gamma^3\right)\gamma^1 \nonumber \\
    & & +\left(\cos\theta\sin\frac{\phi}{2}\cos\frac{\phi}{2}\gamma^1\gamma^2-\cos\theta\sin^2\frac{\phi}{2}-\sin\theta\sin\frac{\phi}{2}\cos\frac{\phi}{2}\gamma^2\gamma^3-\sin\theta\sin^2\frac{\phi}{2}\gamma^1\gamma^3\right)\gamma^1 \nonumber \\
    & & =(\cos\theta\cos\phi+\cos\theta\sin\phi\,\gamma^1\gamma^2+\sin\theta\,\gamma^3\gamma^1)\gamma^1=\cos\theta\cos\phi\,\gamma^1+\cos\theta\sin\phi\,\gamma^2-\sin\theta\,\gamma^3=\gamma^\theta={\bm\gamma}\cdot{\bf n}_\theta,
    \label{gamma1}
\end{eqnarray}
\begin{eqnarray}
    & & \tilde{\gamma}^2=U\gamma^2U^{-1}=U\gamma^2\exp\left(-\frac{1}{2}\theta\gamma^3\gamma^1\right)\exp\left(-\frac{1}{2}\phi\gamma^1\gamma^2\right)=U\exp\left(-\frac{1}{2}\theta\gamma^3\gamma^1\right)\exp\left(\frac{1}{2}\phi\gamma^1\gamma^2\right)\gamma^2 \nonumber \\
    & & =\exp(\phi\gamma^1\gamma^2)\gamma^2=(\cos\phi+\sin\phi\,\gamma^1\gamma^2)\gamma^2=-\sin\phi\,\gamma^1+\cos\phi\,\gamma^2=\gamma^\phi={\bm\gamma}\cdot{\bf n}_\phi,
    \label{gamma2}
\end{eqnarray}
where ${\bm\gamma}$ is the vector composed of $\gamma^1$, $\gamma^2$, and $\gamma^3$, and ${\bf n}_i=(\partial{\bf r}/\partial x^i)/|\partial{\bf r}/\partial x^i|$ are the three orthogonal unit vectors in spherical coordinates:
\begin{eqnarray}
    & & {\bf n}_r=\frac{\partial{\bf r}/\partial r}{|\partial{\bf r}/\partial r|}=\langle\sin\theta\cos\phi,\sin\theta\sin\phi,\cos\theta\rangle={\bf r}/r, \quad {\bf n}_\theta=\frac{\partial{\bf r}/\partial\theta}{|\partial{\bf r}/\partial\theta|}=\langle\cos\theta\cos\phi,\cos\theta\sin\phi,-\sin\theta\rangle \nonumber \\
    & & {\bf n}_\phi=\frac{\partial{\bf r}/\partial\phi}{|\partial{\bf r}/\partial\phi|}=\langle-\sin\phi,\cos\phi,0\rangle. \nonumber
\end{eqnarray}
Consequently, the operator $U$ rotates spatial the Dirac matrices, aligning the coordinate axes with the curvilinear coordinates $r,\theta,\phi$.
The temporal Dirac matrix $\gamma^0$ and the chirality matrix $\gamma^5$ commute with $U$ and therefore do not change under this rotation: $\tilde{\gamma}^0=\gamma^0$, $\tilde{\gamma}^5=\gamma^5$.
The rotated Dirac matrices satisfy the anticommutation relation (\ref{matrixL}) in the Minkowski spacetime:
\[
\{\tilde{\gamma}^\mu,\tilde{\gamma}^\nu\}=2\eta^{\mu\nu}.
\]

The angular part of the Dirac operator (\ref{Diracoperator}) acting on $\psi$ is
\begin{equation}
    D_\Omega\psi=\frac{1}{r}\gamma^3\psi+\frac{1}{r}\gamma^1\left(\partial_\theta+\frac{1}{2}\cot\theta\right)\psi+\frac{1}{r\sin\theta}\gamma^2\partial_\phi\psi,
    \label{angularpart}
\end{equation}
where the non-derivative terms $1/r$ and $(1/2r)\cot\theta$ are generated by the spin connection.
The product
\begin{eqnarray}
    & & U\gamma^1\partial_\theta U^{-1}=U\gamma^1\partial_\theta\left[\exp\left(-\frac{1}{2}\theta\gamma^3\gamma^1\right)\exp\left(-\frac{1}{2}\phi\gamma^1\gamma^2\right)\right]=U\gamma^1\left(-\frac{1}{2}\gamma^3\gamma^1\right)\exp\left(-\frac{1}{2}\theta\gamma^3\gamma^1\right)\exp\left(-\frac{1}{2}\phi\gamma^1\gamma^2\right) \nonumber \\
    & & =-\frac{1}{2}U\gamma^3\exp\left(-\frac{1}{2}\theta\gamma^3\gamma^1\right)\exp\left(-\frac{1}{2}\phi\gamma^1\gamma^2\right)=-\frac{1}{2}U\exp\left(\frac{1}{2}\theta\gamma^3\gamma^1\right)\exp\left(-\frac{1}{2}\phi\gamma^1\gamma^2\right)\gamma^3 \nonumber \\
    & & =-\frac{1}{2}\exp\left(\frac{1}{2}\phi\gamma^1\gamma^2\right)\exp(\theta\gamma^3\gamma^1)\exp\left(-\frac{1}{2}\phi\gamma^1\gamma^2\right)\gamma^3=-\frac{1}{2}(\sin\theta\cos\phi\,\gamma^1+\sin\theta\sin\phi\,\gamma^2+\cos\theta\,\gamma^3)=-\frac{1}{2}\gamma^r
    \label{partheta}
\end{eqnarray}
is similar to the transformation of $\gamma^3$ (\ref{gamma3}).
Therefore, under the rotation generated by $U$, the operator $\gamma^1\partial_\theta$ acting on a spinor $\psi$ transforms according to
\begin{eqnarray}
    & & \gamma^1\partial_\theta\psi=\gamma^1\partial_\theta(U^{-1}U\psi)=\gamma^1\partial_\theta(U^{-1})U\psi+\gamma^1 U^{-1}\partial_\theta(U\psi)=U^{-1}[U\gamma^1\partial_\theta(U^{-1})U\psi+U\gamma^1 U^{-1}\partial_\theta(U\psi)] \nonumber \\
    & & =U^{-1}\left[-\frac{1}{2}\gamma^r\tilde{\psi}+\gamma^\theta\partial_\theta\tilde{\psi}\right],
    \label{rotpartheta}
\end{eqnarray}
using (\ref{gamma1}) and (\ref{partheta}) in the last transition.
A similar calculation gives the product
\begin{eqnarray}
    & & U\gamma^2\partial_\phi U^{-1}=U\gamma^2\partial_\phi\left[\exp\left(-\frac{1}{2}\theta\gamma^3\gamma^1\right)\exp\left(-\frac{1}{2}\phi\gamma^1\gamma^2\right)\right]=U\gamma^2\exp\left(-\frac{1}{2}\theta\gamma^3\gamma^1\right)\left(-\frac{1}{2}\gamma^1\gamma^2\right)\exp\left(-\frac{1}{2}\phi\gamma^1\gamma^2\right) \nonumber \\
    & & =U\exp\left(-\frac{1}{2}\theta\gamma^3\gamma^1\right)\gamma^2\left(-\frac{1}{2}\gamma^1\gamma^2\right)\exp\left(-\frac{1}{2}\phi\gamma^1\gamma^2\right)=-\frac{1}{2}\exp\left(\frac{1}{2}\phi\gamma^1\gamma^2\right)\gamma^1\exp\left(-\frac{1}{2}\phi\gamma^1\gamma^2\right)= \nonumber \\
    & & =-\frac{1}{2}\exp\left(\frac{1}{2}\phi\gamma^1\gamma^2\right)\exp\left(\frac{1}{2}\phi\gamma^1\gamma^2\right)\gamma^1=-\frac{1}{2}\exp(\phi\gamma^1\gamma^2)\gamma^1=-\frac{1}{2}(\cos\phi+\sin\phi\,\gamma^1\gamma^2)\gamma^1 \nonumber \\
    & & =-\frac{1}{2}(\cos\phi\gamma^1+\sin\phi\,\gamma^2).
    \label{parphi}
\end{eqnarray}
Therefore, under the rotation generated by $U$, the operator $\gamma^2\partial_\phi$ acting on $\psi$ transforms according to
\begin{eqnarray}
    & & \gamma^2\partial_\phi\psi=\gamma^2\partial_\phi(U^{-1}U\psi)=\gamma^2\partial_\phi(U^{-1})U\psi+\gamma^2 U^{-1}\partial_\phi(U\psi)=U^{-1}[U\gamma^2\partial_\phi(U^{-1})U\psi+U\gamma^2 U^{-1}\partial_\phi(U\psi)] \nonumber \\
    & & =U^{-1}\left[-\frac{1}{2}(\cos\phi\,\gamma^1+\sin\phi\,\gamma^2)\tilde{\psi}+\gamma^\phi\partial_\phi\tilde{\psi}\right],
    \label{rotparphi}
\end{eqnarray}
using (\ref{gamma2}) and (\ref{parphi}) in the last transition.

Because of (\ref{rotpartheta}) and (\ref{rotparphi}), the angular part (\ref{angularpart}) of the Dirac operator acting on $\psi$ transforms according to
\begin{eqnarray}
    & & D_\Omega\psi=\frac{1}{r}\gamma^3\psi+\frac{1}{r}\gamma^1\left(\partial_\theta\psi+\frac{1}{2}\cot\theta\,\psi\right)+\frac{1}{r\sin\theta}\gamma^2\partial_\phi\psi \nonumber \\
    & & =\gamma^3 U^{-1}\frac{1}{r}\tilde{\psi}+U^{-1}\left[-\frac{1}{2r}\gamma^r\tilde{\psi}+\frac{1}{r}\gamma^\theta\partial_\theta\tilde{\psi}\right]+\gamma^1 U^{-1}\frac{1}{2r}\cot\theta\,\tilde{\psi} \nonumber \\
    & & +U^{-1}\left[-\frac{1}{2r\sin\theta}(\cos\phi\,\gamma^1+\sin\phi\,\gamma^2)\tilde{\psi}+\frac{1}{r\sin\theta}\gamma^\phi\partial_\phi\tilde{\psi}\right], \nonumber
\end{eqnarray}
using $\psi=U^{-1}\tilde{\psi}$.
The terms without derivatives $\partial_\theta$ and $\partial_\phi$ give
\begin{eqnarray}
    & & U^{-1}U\gamma^3 U^{-1}\frac{1}{r}\tilde{\psi}+U^{-1}\left[-\frac{1}{2r}\gamma^r\tilde{\psi}\right]+U^{-1}U\gamma^1 U^{-1}\frac{1}{2r}\cot\theta\,\tilde{\psi}+U^{-1}\left[-\frac{1}{2r\sin\theta}(\cos\phi\,\gamma^1+\sin\phi\,\gamma^2)\tilde{\psi}\right] \nonumber \\
    & & =U^{-1}\left[\gamma^r-\frac{1}{2}\gamma^r+\frac{1}{2}\gamma^\theta\cot\theta-\frac{1}{2\sin\theta}(\cos\phi\,\gamma^1+\sin\phi\,\gamma^2)\right]\frac{1}{r}\tilde{\psi} \nonumber \\
    & & =\frac{1}{2r}U^{-1}[\sin\theta\cos\phi\,\gamma^1+\sin\theta\sin\phi\,\gamma^2+\cos\theta\,\gamma^3+\cot\theta(\cos\theta\cos\phi\,\gamma^1+\cos\theta\sin\phi\,\gamma^2-\sin\theta\,\gamma^3) \nonumber \\
    & & -(\cos\phi\,\gamma^1+\sin\phi\,\gamma^2)/\sin\theta]\tilde{\psi}=0. \nonumber
\end{eqnarray}
Therefore, the angular part of the Dirac operator acting on the spinor $\tilde{\psi}$ retains only the terms with derivatives $\partial_\theta$ and $\partial_\phi$:
\begin{equation}
    D_\Omega\psi=U^{-1}\left[\frac{1}{r}\gamma^\theta\partial_\theta\tilde{\psi}+\frac{1}{r\sin\theta}\gamma^\phi\partial_\phi\tilde{\psi}\right]=U^{-1}\tilde{D}_\Omega\tilde{\psi}.
    \label{retain}
\end{equation}
The rotation generated by the unitary spin-rotation operator $U$ (\ref{unitary}) subtracts the spin-connection terms $1/r$ and $(1/2r)\cot\theta$ from the angular Dirac operator and replaces the constant Dirac matrices $\gamma^i$ with the rotated, position-dependent Dirac matrices $\tilde{\gamma}^i$ \cite{rotations}:\footnote{
The spin-connection terms could also be removed from $\tilde{D}_\Omega$ by rescaling the spinor: $\psi=\tilde{\psi}/r(\sin\theta)^{1/2}$.
However, this rescaling would add the dependence on $\theta$ of the nonlinear term in the Dirac equation, which could not be reduced to a system of radial equations.
}
\begin{equation}
    \tilde{D}_\Omega=\frac{1}{r}\gamma^\theta\partial_\theta+\frac{1}{r\sin\theta}\gamma^\phi\partial_\phi.
    \label{angularoperator}
\end{equation}
This operator is related to the angular momentum of the spinor.\footnote{
Using $\gamma^3\gamma^1=-i\Sigma_2$ and $\gamma^2\gamma^3=-i\Sigma_1$, the rotated matrices satisfy $\gamma^r\gamma^\theta=-i\Sigma_\phi$ and $\gamma^\phi\gamma^r=-i\Sigma_\theta$, which is equivalent to $\gamma^\theta=i\gamma^r\Sigma_\phi$ and $\gamma^\phi=-i\gamma^r\Sigma_\theta$.
Therefore, the angular Dirac operator (\ref{angularoperator}) is
\[
    \tilde{D}_\Omega=\frac{i}{r}\gamma^r\Sigma_\phi\partial_\theta-\frac{i}{r\sin\theta}\gamma^r\Sigma_\theta\partial_\phi.
\]
The orbital angular momentum operator ${\bf L}={\bf r}\times{\bf p}$, where ${\bf p}$ is the momentum operator has rectangular components in terms of the spherical coordinates \cite{Dirac}:
\[
    {\bf L}=\langle i\hbar(\sin\phi\,\partial_\theta+\cot\theta\cos\phi\,\partial_\phi),i\hbar(-\cos\phi\,\partial_\theta+\cot\theta\sin\phi\,\partial_\phi),-i\hbar\partial_\phi\rangle.
\]
so its spherical components are
\[
    L_\theta={\bf L}\cdot{\bf n}_\theta=\frac{i\hbar}{\sin\theta}\partial_\phi,\quad L_\phi={\bf L}\cdot{\bf n}_\phi=-i\hbar\partial_\theta,\quad L_r={\bf L}\cdot{\bf n}_r=0.
\]
Therefore, the angular Dirac operator becomes
\[
    \tilde{D}_\Omega=-\frac{1}{\hbar r}\gamma^r(\Sigma_\theta L_\theta+\Sigma_\phi L_\phi)=-\frac{1}{\hbar r}\gamma^r({\bm\Sigma}\cdot{\bf L}),
\]
where ${\bm\Sigma}$ is the vector formed from the spin matrices.
This operator can be written in terms of the spin-orbit operator \cite{RQM,QFT}
\[
K=-\beta({\bm\Sigma}\cdot{\bf L}+\hbar)
\]
as
\begin{equation}
    \tilde{D}_\Omega=\frac{1}{\hbar r}\gamma^r(\beta K+\hbar).
    \label{spinorbit}
\end{equation}
}\\

{\bf 9. Rotation of the curvature and torsion terms in the Dirac equation}\\ \\
The Dirac operator (\ref{Diracoperator}) acting on $\psi$ is equivalent to
\begin{eqnarray}
    & & D\psi=e^{-\nu/2}\gamma^0\left(\partial_{ct}+\frac{1}{4}\dot{\lambda}\right)\psi+e^{-\lambda/2}\gamma^3\left(\partial_r+\frac{1}{r}+\frac{1}{4}\nu'\right)\psi-\frac{1}{r}\gamma^3\psi+D_\Omega\psi,
    \label{Diracoperatorshort}
\end{eqnarray}
where $D_\Omega\psi$ is the angular part (\ref{angularpart}).
The operators $\partial_{ct}$ and $\partial_r$ commute with $U^{-1}$ because $U$ does not depend on $t$ and $r$.
Consequently, the Dirac operator (\ref{Diracoperatorshort}) acting on $\psi$ transforms according to
\begin{eqnarray}
    & & D\psi=e^{-\nu/2}\gamma^0 U^{-1}\left(\partial_{ct}+\frac{1}{4}\dot{\lambda}\right)\tilde{\psi}+e^{-\lambda/2}\gamma^3 U^{-1}\left(\partial_r+\frac{1}{r}+\frac{1}{4}\nu'\right)\tilde{\psi}-\frac{1}{r}\gamma^3 U^{-1}\tilde{\psi}+U^{-1}\tilde{D}_\Omega\tilde{\psi}, \nonumber \\
    & & =e^{-\nu/2}U^{-1}\gamma^0\left(\partial_{ct}+\frac{1}{4}\dot{\lambda}\right)\tilde{\psi}+e^{-\lambda/2}U^{-1}(U\gamma^3 U^{-1})\left(\partial_r+\frac{1}{r}+\frac{1}{4}\nu'\right)\tilde{\psi}-\frac{1}{r}U^{-1}(U\gamma^3 U^{-1})\tilde{\psi}+U^{-1}\tilde{D}_\Omega\tilde{\psi} \nonumber \\
    & & =U^{-1}\left[e^{-\nu/2}\gamma^0\left(\partial_{ct}+\frac{1}{4}\dot{\lambda}\right)+e^{-\lambda/2}\gamma^r\left(\partial_r+\frac{1}{r}+\frac{1}{4}\nu'\right)-\frac{1}{r}\gamma^r+\tilde{D}_\Omega\right]\tilde{\psi}=U^{-1}\tilde{D}\tilde{\psi}.
\end{eqnarray}
using (\ref{retain}) for the last term.
Therefore, the rotated Dirac operator acting on a rotated spinor $\tilde{\psi}$ is
\begin{equation}
    \tilde{D}\tilde{\psi}=e^{-\nu/2}\gamma^0\left(\partial_{ct}+\frac{1}{4}\dot{\lambda}\right)\tilde{\psi}+e^{-\lambda/2}\gamma^r\left(\partial_r+\frac{1}{r}+\frac{1}{4}\nu'\right)\tilde{\psi}-\frac{1}{r}\gamma^r\tilde{\psi}+\frac{1}{r}\gamma^\theta\partial_\theta\tilde{\psi}+\frac{1}{r\sin\theta}\gamma^\phi\partial_\phi\tilde{\psi}.
    \label{Diracoperatorrotated}
\end{equation}

Under this rotation, the axial current is invariant:
\begin{eqnarray}
    j^{5\mu}=\overline{\psi}\gamma^\mu\gamma^5\psi=\overline{\psi}U^{-1}(U\gamma^\mu U^{-1})(U\gamma^5 U^{-1})U\psi=\overline{\tilde{\psi}}\tilde{\gamma}^\mu\tilde{\gamma}^5\tilde{\psi}=\tilde{j}^{5\mu}.
\end{eqnarray}
Therefore, the cubic torsion term in the Dirac equation acting on a spinor $\psi$ transforms according to
\begin{eqnarray}
    j^{5\mu}\gamma_\mu\gamma^5\psi=\tilde{j}^{5\mu}U^{-1}(U\gamma_\mu U^{-1})(U\gamma^5 U^{-1})U\psi=U^{-1}\tilde{j}^{5\mu}\tilde{\gamma}_\mu\tilde{\gamma}^5\tilde{\psi}.
\end{eqnarray}
Consequently, if the Dirac equation (\ref{Diracspherical}) is satisfied for $\psi$:
\[
    i\hbar D\psi=mc\psi-\mathcal{K}j^{5\mu}\gamma_\mu\gamma^5\psi,\quad i\hbar U^{-1}\tilde{D}\tilde{\psi}=mcU^{-1}\tilde{\psi}-\mathcal{K}U^{-1}\tilde{j}^{5\mu}\tilde{\gamma}_\mu\tilde{\gamma}^5\tilde{\psi},
\]
then it has the same form for $\tilde{\psi}$:
\[
    i\hbar\tilde{D}\tilde{\psi}=mc\tilde{\psi}-\mathcal{K}\tilde{j}^{5\mu}\tilde{\gamma}_\mu\tilde{\gamma}^5\tilde{\psi}.
\]
Putting the rotated Dirac operator (\ref{Diracoperatorrotated}) into this equation gives the Dirac equation in a centrally symmetric gravitational field with curvature and torsion:
\begin{equation}
    i\hbar\left[e^{-\nu/2}\gamma^0\left(\partial_{ct}+\frac{1}{4}\dot{\lambda}\right)+e^{-\lambda/2}\gamma^r\left(\partial_r+\frac{1}{r}+\frac{1}{4}\nu'\right)-\gamma^r\frac{1}{r}+\tilde{D}_\Omega\right]\tilde{\psi}=(mc-\mathcal{K}\tilde{j}^{5\mu}\tilde{\gamma}_\mu\tilde{\gamma}^5)\tilde{\psi}.
    \label{Diraccentral}
\end{equation}

For a stationary state, the functions $\nu$ and $\lambda$ in (\ref{centralmetric}) depend only on $r$, so $\dot{\lambda}=0$ and, according to (\ref{spinorconnection}), $\Gamma_r=0$.
The spinor $\tilde{\psi}$ in this state is proportional to $e^{-iEt/\hbar}$, where $E$ is its energy.
The Dirac equation (\ref{Diraccentral}) therefore becomes
\begin{eqnarray}
    & & e^{-\nu/2}E\tilde{\psi}=-i\hbar c\gamma^0\left[e^{-\lambda/2}\gamma^r\left(\partial_r+\frac{1}{r}+\frac{1}{4}\nu'\right)-\gamma^r\frac{1}{r}+\tilde{D}_\Omega\right]\tilde{\psi}+\gamma^0(mc^2-\mathcal{K}c\tilde{j}^{5\mu}\tilde{\gamma}_\mu\tilde{\gamma}^5)\tilde{\psi} \nonumber \\
    & & =-i\hbar c\left[e^{-\lambda/2}\alpha^r\left(\partial_r+\frac{1}{r}+\frac{1}{4}\nu'\right)-\alpha^r\frac{1}{r}+\beta\tilde{D}_\Omega\right]\tilde{\psi}+\beta(mc^2-\mathcal{K}c\tilde{j}^{5\mu}\tilde{\gamma}_\mu\tilde{\gamma}^5)\tilde{\psi},
    \label{Diracstationary}
\end{eqnarray}
where $\nu$ and $\lambda$ are functions of $r$.\\

{\bf 10. The Dirac equation for a spherical spinor with torsion}\\ \\
In the Dirac representation \cite{Dirac,RQM,QFT},       $\gamma^0=\beta=\left( \begin{array}{cc}
    I_2 & 0 \\
    0 & -I_2 \end{array} \right), \gamma^\mu=\left( \begin{array}{cc}
    0 & \sigma_\mu \\
    -\sigma_\mu & 0 \end{array} \right), \alpha^\mu=\beta\gamma^\mu=\left( \begin{array}{cc}
    0 & \sigma_\mu \\
    \sigma_\mu & 0 \end{array} \right)$, where $I_2$ is the two-dimensional unit matrix, $\mu=1,2,3$, and $\sigma_\mu$ are the Pauli matrices: $\sigma_1=\left( \begin{array}{cc}
    0 & 1 \\
    1 & 0 \end{array} \right)$, $\sigma_2=\left( \begin{array}{cc}
    0 & -i \\
    i & 0 \end{array} \right)$, $\sigma_3=\left( \begin{array}{cc}
    1 & 0 \\
    0 & -1 \end{array} \right)$.
The spin matrices are $\Sigma_\mu=\left( \begin{array}{cc}
    \sigma_\mu & 0 \\
    0 & \sigma_\mu \end{array} \right)$.
The chirality matrix is $\gamma^5=\left( \begin{array}{cc}
    0 & I_2 \\
    I_2 & 0 \end{array} \right)$.

A unitary operator, analogous to (\ref{unitary}):
\begin{eqnarray}
    & & U=\exp\left(-\frac{i}{2}\phi\sigma_3\right)\exp\left(-\frac{i}{2}\theta\sigma_2\right),\quad U^\dagger=U^{-1}=\exp\left(\frac{i}{2}\theta\sigma_2\right)\exp\left(\frac{i}{2}\phi\sigma_3\right)
\end{eqnarray}
generates a rotation of two-component Pauli spinors:
\begin{eqnarray}
    & & \tilde{\sigma}_3=U\sigma_3 U^{-1}=\sin\theta\cos\phi\,\sigma_1+\sin\theta\sin\phi\,\sigma_2+\cos\theta\,\sigma_3=\sigma_r={\bm\sigma}\cdot{\bf n}_r, \nonumber \\
    & & \tilde{\sigma}_1=U\sigma_1 U^{-1}=\cos\theta\cos\phi\,\sigma_1+\cos\theta\sin\phi\,\sigma_2-\sin\theta\,\sigma_3=\sigma_\theta={\bm\sigma}\cdot{\bf n}_\theta, \nonumber \\
    & & \tilde{\sigma}_2=U\sigma_2 U^{-1}=-\sin\phi\,\sigma_1+\cos\phi\,\sigma_2=\sigma_\phi={\bm\sigma}\cdot{\bf n}_\phi,
\end{eqnarray}
analogously to (\ref{gamma3}), (\ref{gamma1}), and (\ref{gamma2}), where ${\bm\sigma}$ is the vector formed from the Pauli matrices.
Consequently, the rotated Dirac matrices are $\tilde{\gamma}^\mu=\left( \begin{array}{cc}
    0 & \tilde{\sigma}_\mu \\
    -\tilde{\sigma}_\mu & 0 \end{array} \right)$, where the rotated Pauli matrices are
\[
\sigma_r=\left( \begin{array}{cc}
    \cos\theta & \sin\theta\,e^{-i\phi} \\
    \sin\theta\,e^{i\phi}& -\cos\theta \end{array} \right),\quad \sigma_\theta=\left( \begin{array}{cc}
    -\sin\theta & \cos\theta\,e^{-i\phi} \\
    \cos\theta\,e^{i\phi}& \sin\theta \end{array} \right),\quad \sigma_\phi=\left( \begin{array}{cc}
    0 & -ie^{-i\phi} \\
    ie^{i\phi} & 0 \end{array} \right).
\]
Because $\partial_\theta{\bf n}_r={\bf n}_\theta$ and $\partial_\phi{\bf n}_r=\sin\theta\,{\bf n}_\phi$, the Pauli matrices satisfy
\begin{equation}
    \partial_\theta\sigma_r=\sigma_\theta,\quad \partial_\phi\sigma_r=\sin\theta\,\sigma_\phi.
    \label{derivatives}
\end{equation}

We seek a solution of the Dirac equation (\ref{Diracstationary}) in the form of the Wakano spherical spinor \cite{Wakano}:
\begin{equation}
    \tilde{\psi}^{+}=\left( \begin{array}{c}
    g(r)\xi \\
    if(r)({\bm\sigma}\cdot{\bf n})\xi \\ \end{array} \right)\quad\mbox{for}\,E>0,\quad \tilde{\psi}^{-}=\left( \begin{array}{c}
    if(r)({\bm\sigma}\cdot{\bf n})\xi \\
    g(r)\xi \\ \end{array} \right)\quad\mbox{for}\,E<0,
    \label{spinorsolution}
\end{equation}
where $\xi$ is a normalized, constant Pauli spinor ($\xi\dagger\xi=1$), $g(r)$ is a large-component function, $f(r)$ is a small-component function, and ${\bf n}={\bf n}_r$.\footnote{
In the nonrelativistic limit, the ratio $f/g$ is on the order of $p/2mc$.
}
The factor $i$ and the operator ${\bm\sigma}\cdot{\bf n}$ couple 
the orbital angular momentum and spin while ensuring that the upper and lower components of the spinor have opposite parity under spatial inversion.\footnote{
Putting (\ref{spinorbit}) in the Dirac equation (\ref{Diracstationary}) gives
\begin{equation}
    e^{-\nu/2}E\tilde{\psi}=-i\hbar c\left[e^{-\lambda/2}\alpha^r\left(\partial_r+\frac{1}{r}+\frac{1}{4}\nu'\right)+\frac{1}{\hbar r}\alpha^r\beta K\right]\tilde{\psi}+\beta(mc^2-\mathcal{K}c\tilde{j}^{5\mu}\tilde{\gamma}_\mu\tilde{\gamma}^5)\tilde{\psi}.
    \label{DiracK}
\end{equation}
Without curvature and torsion, this equation reduces to the standard form (equivalent to $i\hbar\partial_t\psi=-i\hbar c{\bm\alpha}\cdot{\bm\nabla}\psi+mc^2\beta\psi$) \cite{RQM}:
\begin{eqnarray}
    & & E\tilde{\psi}=\beta mc^2\tilde{\psi}-i\hbar c\alpha^r\left(\partial_r+\frac{1}{r}+\frac{1}{\hbar r}\beta K\right)\tilde{\psi}=\beta mc^2\tilde{\psi}-i\hbar c\alpha^r\left(\partial_r-\frac{1}{\hbar r}({\bm\Sigma}\cdot{\bf L})\right)\tilde{\psi}=\beta mc^2\tilde{\psi}+\alpha^r\left(p_r+\frac{i}{r}({\bm\Sigma}\cdot{\bf L})\right)c\tilde{\psi} \nonumber \\
    & & =\beta mc^2\tilde{\psi}+({\bm\alpha}\cdot{\bf p})c\tilde{\psi}, \nonumber
\end{eqnarray}
where ${\bm\alpha}$ is the vector formed from the matrices $\alpha^\mu$.
The last equality follows from the identity $({\bm\sigma}\cdot{\bf r})({\bm\sigma}\cdot{\bf L})=\sigma_i\sigma_k r_i L_k=(\delta_{ik}+i\epsilon_{ikl}\sigma_l)r_i L_k=i({\bf r}\times{\bf L})\cdot{\bm\sigma}=i[({\bf r}\cdot{\bf p}){\bf r}-r^2{\bf p}]\cdot{\bm\sigma}$.
Replacing ${\bm\sigma}$ with ${\bm\Sigma}$ gives $({\bm\Sigma}\cdot{\bf r})({\bm\Sigma}\cdot{\bf L})=i[({\bf r}\cdot{\bf p}){\bf r}-r^2{\bf p}]\cdot{\bm\Sigma}$, and using ${\bm\Sigma}=\gamma^5{\bm\alpha}$ gives $({\bm\alpha}\cdot{\bf r})({\bm\Sigma}\cdot{\bf L})=i[({\bf r}\cdot{\bf p}){\bf r}-r^2{\bf p}]\cdot{\bm\alpha}$ and therefore $\alpha^r({\bm\Sigma}\cdot{\bf L})=ir[\alpha^r p_r-({\bm\alpha}\cdot{\bf p})]$, which is equivalent to ${\bm\alpha}\cdot{\bf p}=\alpha^r[p_r+i({\bm\Sigma}\cdot{\bf L})/r]$.
}
The constancy of $\xi$ simplifies the dependence on the angles of the nonlinear term in the Dirac equation and corresponds to the ground state of the spinor.

We use (\ref{angularoperator}) and (\ref{derivatives}) to determine the term $\tilde{D}_\Omega\tilde{\psi}$ in (\ref{Diracstationary}) for a positive-energy state $E>0$:
\begin{eqnarray}
    & & \tilde{D}_\Omega\tilde{\psi}^{+}=\frac{1}{r}\left( \begin{array}{cc}
    0 & \sigma_\theta \\
    -\sigma_\theta & 0 \end{array} \right)\partial_\theta\left( \begin{array}{c}
    g(r)\xi \\
    if(r)({\bm\sigma}\cdot{\bf n})\xi \\ \end{array} \right)+\frac{1}{r\sin\theta}\left( \begin{array}{cc}
    0 & \sigma_\phi \\
    -\sigma_\phi & 0 \end{array} \right)\partial_\phi\left( \begin{array}{c}
    g(r)\xi \\
    if(r)({\bm\sigma}\cdot{\bf n})\xi \\ \end{array} \right) \nonumber \\
    & & =\frac{1}{r}\left( \begin{array}{cc}
    0 & \sigma_\theta \\
    -\sigma_\theta & 0 \end{array} \right)\partial_\theta\left( \begin{array}{c}
    g\xi \\
    if\sigma_r\xi \\ \end{array} \right)+\frac{1}{r\sin\theta}\left( \begin{array}{cc}
    0 & \sigma_\phi \\
    -\sigma_\phi & 0 \end{array} \right)\partial_\phi\left( \begin{array}{c}
    g\xi \\
    if\sigma_r\xi \\ \end{array} \right) \nonumber \\
    & & =\frac{1}{r}\left( \begin{array}{cc}
    0 & \sigma_\theta \\
    -\sigma_\theta & 0 \end{array} \right)\left( \begin{array}{c}
    0 \\
    if\sigma_\theta\xi \\ \end{array} \right)+\frac{1}{r\sin\theta}\left( \begin{array}{cc}
    0 & \sigma_\phi \\
    -\sigma_\phi & 0 \end{array} \right)\left( \begin{array}{c}
    0 \\
    if\sin\theta\,\sigma_\phi\xi \\ \end{array} \right)=\frac{2}{r}\left( \begin{array}{c}
    if\xi \\
    0 \\ \end{array} \right).
    \label{angularpositive}
\end{eqnarray}
Similarly, for a negative-energy state $E<0$:
\begin{eqnarray}
    & & \tilde{D}_\Omega\tilde{\psi}^{-}=\frac{2}{r}\left( \begin{array}{c}
    0 \\
    -if\xi \\ \end{array} \right).
    \label{angularnegative}
\end{eqnarray}

We determine the term $\tilde{j}^{5\mu}\tilde{\gamma}_\mu\tilde{\gamma}^5\tilde{\psi}$ in (\ref{Diracstationary}) for $E>0$.
Using (\ref{adjointL}), the temporal component of the axial current (\ref{axialcurrent}) is
\[
    \tilde{j}^{50}=\overline{\tilde{\psi}}\tilde{\gamma}^0\tilde{\gamma}^5\tilde{\psi}=0.
\]
The spatial components are
\begin{eqnarray}
    & & \tilde{j}^{5\mu}=\overline{\tilde{\psi}}\tilde{\gamma}^\mu\tilde{\gamma}^5\tilde{\psi}=[g\xi^\dagger\quad if\xi^\dagger({\bm\sigma}\cdot{\bf n})]\left( \begin{array}{cc}
    \sigma_\mu & 0 \\
    0 & -\sigma_\mu \end{array} \right)\left( \begin{array}{c}
    g\xi \\
    if({\bm\sigma}\cdot{\bf n})\xi \\ \end{array} \right) \nonumber \\
    & & =[g\xi^\dagger\quad if\xi^\dagger({\bm\sigma}\cdot{\bf n})]\left( \begin{array}{c}
    \sigma_\mu g\xi \\
    -\sigma_\mu if(r)({\bm\sigma}\cdot{\bf n})\xi \\ \end{array} \right)=g^2\xi^\dagger\sigma_\mu\xi+f^2\xi^\dagger({\bm\sigma}\cdot{\bf n})\sigma_\mu({\bm\sigma}\cdot{\bf n})\xi \nonumber \\
    & & =g^2\xi^\dagger\sigma_\mu\xi+2f^2\xi^\dagger({\bm\sigma}\cdot{\bf n})\xi n_\mu-f^2\xi^\dagger\sigma_\mu\xi=(g^2-f^2)s_\mu+2f^2({\bf s}\cdot{\bf n})n_\mu, \nonumber
\end{eqnarray}
where
\[
s_\mu=\xi^\dagger\sigma_\mu\xi
\]
corresponds to the expectation value of spin and the identity $({\bm\sigma}\cdot{\bf n})\sigma_\mu({\bm\sigma}\cdot{\bf n})+\sigma_\mu=2({\bm\sigma}\cdot{\bf n})n_\mu$ was used.
Therefore, the axial current is
\[
{\bf j}^5=(g^2-f^2){\bf s}+2f^2({\bf s}\cdot{\bf n}){\bf n}.
\]

The contraction of the axial current $\tilde{j}^{5\mu}$ with $\tilde{\gamma}_\mu\tilde{\gamma}^5\tilde{\psi}$ gives
\begin{eqnarray}
    & & \tilde{j}^{5\mu}\tilde{\gamma}_\mu\tilde{\gamma}^5\tilde{\psi}=-{\bf j}^5
    \cdot\tilde{\bm\gamma}\tilde{\gamma}^5\tilde{\psi}=\left( \begin{array}{cc}
    -{\bf j}^5\cdot{\bm\sigma} & 0 \\
    0 & {\bf j}^5\cdot{\bm\sigma} \end{array} \right)\left( \begin{array}{c}
    g\xi \\
    if({\bm\sigma}\cdot{\bf n})\xi \\ \end{array} \right)=\left( \begin{array}{c}
    -g({\bf j}^5\cdot{\bm\sigma})\xi \\
    if({\bf j}^5\cdot{\bm\sigma})({\bm\sigma}\cdot{\bf n})\xi \\ \end{array} \right) \nonumber \\
    & & =\left( \begin{array}{c}
    -g(g^2-f^2)({\bf s}\cdot{\bm\sigma})\xi-2gf^2({\bf s}\cdot{\bf n})({\bm\sigma}\cdot{\bf n})\xi \\
    if(g^2-f^2)({\bf s}\cdot{\bm\sigma})({\bm\sigma}\cdot{\bf n})\xi+2if^3({\bf s}\cdot{\bf n})({\bm\sigma}\cdot{\bf n})({\bm\sigma}\cdot{\bf n})\xi \end{array} \right)= \nonumber \\
    & & =\left( \begin{array}{c}
    -g(g^2-f^2)\xi-2gf^2({\bf s}\cdot{\bf n})({\bm\sigma}\cdot{\bf n})\xi \\
    if(g^2-f^2)[2{\bf s}\cdot{\bf n}-({\bm\sigma}\cdot{\bf n})({\bf s}\cdot{\bm\sigma})]\xi+2if^3({\bf s}\cdot{\bf n})\xi \end{array} \right)=\left( \begin{array}{c}
    -g(g^2-f^2)\xi-2gf^2({\bf s}\cdot{\bf n})({\bm\sigma}\cdot{\bf n})\xi \\
    2ifg^2({\bf s}\cdot{\bf n})\xi-if(g^2-f^2)({\bm\sigma}\cdot{\bf n})\xi \end{array} \right) \nonumber \\
    & & =-(g^2-f^2)\left( \begin{array}{c}
    g\xi \\
    if({\bm\sigma}\cdot{\bf n})\xi \\ \end{array} \right)+2igf({\bf s}\cdot{\bf n})\left( \begin{array}{c}
    if({\bm\sigma}\cdot{\bf n})\xi \\ g\xi \end{array} \right)=-(g^2-f^2)\tilde{\psi}+2igf({\bf s}\cdot{\bf n})\tilde{\gamma}^5\tilde{\psi},
    \label{axialpositive}
\end{eqnarray}
using the identities $({\bm\sigma}\cdot{\bf n})({\bm\sigma}\cdot{\bf n})=I_2$, $({\bf s}\cdot{\bm\sigma})({\bm\sigma}\cdot{\bf n})=2{\bf s}\cdot{\bf n}-({\bm\sigma}\cdot{\bf n})({\bf s}\cdot{\bm\sigma})$, and $({\bf s}\cdot{\bm\sigma})\xi=\xi$.
Similarly, for $E<0$:
\begin{equation}
    \tilde{j}^{5\mu}\tilde{\gamma}_\mu\tilde{\gamma}^5\tilde{\psi}=(g^2-f^2)\left( \begin{array}{c}
    if({\bm\sigma}\cdot{\bf n})\xi \\ g\xi \end{array} \right)-2igf({\bf s}\cdot{\bf n})\left( \begin{array}{c}
    g\xi \\
    if({\bm\sigma}\cdot{\bf n})\xi \\ \end{array} \right)=(g^2-f^2)\tilde{\psi}-2igf({\bf s}\cdot{\bf n})\tilde{\gamma}^5\tilde{\psi}.
    \label{axialnegative}
\end{equation}

Putting the angular term (\ref{angularpositive}) and the axial term (\ref{axialpositive}) in the Dirac equation (\ref{Diracstationary}) gives for $E>0$:
\begin{eqnarray}
    & & e^{-\nu/2}E\left( \begin{array}{c}
    g\xi \\
    if\sigma_r\xi \\ \end{array} \right)=-i\hbar c\left[e^{-\lambda/2}\left(\partial_r+\frac{1}{r}+\frac{1}{4}\nu'\right)\alpha^r\left( \begin{array}{c}
    g\xi \\
    if\sigma_r\xi \\ \end{array} \right)-\frac{1}{r}\alpha^r\left( \begin{array}{c}
    g\xi \\
    if\sigma_r\xi \\ \end{array} \right)+\frac{2}{r}\beta\left( \begin{array}{c}
    if\xi \\
    0 \\ \end{array} \right)\right] \nonumber \\
    & & +mc^2\beta\left( \begin{array}{c}
    g\xi \\
    if\sigma_r\xi \\ \end{array} \right)+\mathcal{K}c(g^2-f^2)\beta\left( \begin{array}{c}
    g\xi \\
    if({\bm\sigma}\cdot{\bf n})\xi \\ \end{array} \right)-2i\mathcal{K}cgf({\bf s}\cdot{\bf n})\beta\left( \begin{array}{c}
    if({\bm\sigma}\cdot{\bf n})\xi \\ g\xi \end{array} \right).
\end{eqnarray}
Using $\alpha^r=\left( \begin{array}{cc}
    0 & \sigma_r \\
    \sigma_r & 0 \end{array} \right)$ and $(\sigma_r)^2=I_2$ gives for the upper component:
\[
    e^{-\nu/2}Eg\xi=\hbar c\left[e^{-\lambda/2}\left(\partial_r+\frac{1}{r}+\frac{1}{4}\nu'\right)f\xi-\frac{1}{r}f\xi+\frac{2}{r}f\xi\right]+mc^2 g\xi+\mathcal{K}c(g^2-f^2)g\xi+2\mathcal{K}cgf^2({\bf s}\cdot{\bf n})({\bm\sigma}\cdot{\bf n})\xi.
\]
Because of spherical symmetry, the product $({\bf s}\cdot{\bf n})({\bm\sigma}\cdot{\bf n})\xi$ can be averaged over all directions, using $\overline{n_\mu n_\nu}=(1/3)\delta_{\mu\nu}$, which gives $(1/3)({\bf s}\cdot{\bm\sigma})\xi=(1/3)\xi$.
The equation for the upper component therefore simplifies to
\begin{equation}
    (e^{-\nu/2}E-mc^2)g=\hbar c\left[e^{-\lambda/2}\left(\partial_r+\frac{1}{r}+\frac{1}{4}\nu'\right)f+\frac{1}{r}f\right]+\frac{1}{3}\mathcal{K}cg(3g^2-f^2).
    \label{upperpositive}
\end{equation}
A similar calculation for the lower component gives
\[
    e^{-\nu/2}Ef\sigma_r\xi=-\hbar c\left[e^{-\lambda/2}\left(\partial_r+\frac{1}{r}+\frac{1}{4}\nu'\right)g\sigma_r\xi-\frac{1}{r}g\sigma_r\xi\right]-mc^2 f\sigma_r\xi-\mathcal{K}c(g^2-f^2)f({\bm\sigma}\cdot{\bf n})\xi+2\mathcal{K}cfg^2({\bf s}\cdot{\bf n})\xi,
\]
which, when multiplied by $\sigma_r$ from the left, simplifies to
\begin{equation}
    (e^{-\nu/2}E+mc^2)f=-\hbar c\left[e^{-\lambda/2}\left(\partial_r+\frac{1}{r}+\frac{1}{4}\nu'\right)g-\frac{1}{r}g\right]+\frac{1}{3}\mathcal{K}cf(3f^2-g^2).
    \label{lowerpositive}
\end{equation}

Putting the angular term (\ref{angularnegative}) and the axial term (\ref{axialnegative}) in the Dirac equation (\ref{Diracstationary}) gives for $E<0$:
\begin{eqnarray}
    & & e^{-\nu/2}E\left( \begin{array}{c}
    if\sigma_r\xi \\
    g\xi \\ \end{array} \right)=-i\hbar c\left[e^{-\lambda/2}\left(\partial_r+\frac{1}{r}+\frac{1}{4}\nu'\right)\alpha^r\left( \begin{array}{c}
    if\sigma_r\xi \\
    g\xi \\ \end{array} \right)-\frac{1}{r}\alpha^r\left( \begin{array}{c}
    if\sigma_r\xi \\
    g\xi \\ \end{array} \right)+\frac{2}{r}\beta\left( \begin{array}{c}
    0 \\
    -if\xi \\ \end{array} \right)\right] \nonumber \\
    & & +mc^2\beta\left( \begin{array}{c}
    if\sigma_r\xi \\
    g\xi \\ \end{array} \right)-\mathcal{K}c(g^2-f^2)\beta\left( \begin{array}{c}
    if({\bm\sigma}\cdot{\bf n})\xi \\ g\xi \end{array} \right)+2i\mathcal{K}cgf({\bf s}\cdot{\bf n})\beta\left( \begin{array}{c}
    g\xi \\
    if({\bm\sigma}\cdot{\bf n})\xi \\ \end{array} \right).
\end{eqnarray}
Similar calculations give the equation for the upper component:
\begin{equation}
    (e^{-\nu/2}E-mc^2)f=-\hbar c\left[e^{-\lambda/2}\left(\partial_r+\frac{1}{r}+\frac{1}{4}\nu'\right)g-\frac{1}{r}g\right]+\frac{1}{3}\mathcal{K}cf(3f^2-g^2),
    \label{uppernegative}
\end{equation}
and the lower component:
\begin{equation}
    (e^{-\nu/2}E+mc^2)g=\hbar c\left[e^{-\lambda/2}\left(\partial_r+\frac{1}{r}+\frac{1}{4}\nu'\right)f+\frac{1}{r}f\right]+\frac{1}{3}\mathcal{K}cg(3g^2-f^2).
    \label{lowernegative}
\end{equation}
Equations (\ref{upperpositive}) and (\ref{lowerpositive}) for the positive-energy state and equations (\ref{uppernegative}) and (\ref{lowernegative}) for the negative-energy state are nonlinear first-order ordinary differential equations giving functions $g(r)$ and $f(r)$ if the functions $\nu(r)$ and $\lambda(r)$ are determined by an external field.\footnote{
Comparing (\ref{DiracK}) with (\ref{upperpositive}) or (\ref{lowerpositive}) shows that the eigenvalue of $K$ for the spherical spinor (\ref{spinorsolution}) with $E>0$ is equal to $-\hbar$.
Comparing (\ref{DiracK}) with (\ref{uppernegative}) or (\ref{lowernegative}) shows that the eigenvalue of $K$ for the spherical spinor (\ref{spinorsolution}) with $E<0$ is equal to $\hbar$.
}
If the functions $\nu(r)$ and $\lambda(r)$ represent the gravitational field of the spinor, they are determined by the Einstein field equations, shown in Section 13, with the energy--momentum tensor for the spherical spinor, derived in Section 12.\footnote{
In flat spacetime, without curvature and torsion, $\nu=\lambda=0$ and $\mathcal{K}=0$.
Equations (\ref{upperpositive}) and (\ref{lowerpositive}) reduce to those for a free spinor:
\[
    k_1 g=df/dr+2f/r,\quad k_2 f=-dg/dr,
\]
and equations (\ref{uppernegative}) and (\ref{lowernegative}) reduce to the same equations with $f$ replaced with $-f$ and $E$ replaced with $-E$, where
$k_1=(E-mc^2)/\hbar c$, $k_2=(E+mc^2)/\hbar c$, $k=\sqrt{k_1 k_2}=p/\hbar$.
Combining these two equations gives
\[
r^2 d^2g/dr^2 + 2r\,dg/dr + (kr)^2 g = 0,
\]
whose regular solution is $g(r)=\mbox{const}\times j_0(kr)$, and therefore $f(r)=\mbox{const}\times\sqrt{k_1/k_2}\,j_1(kr)$, where $j_l(r)$ is the spherical Bessel function of the first kind of order $l$ (corresponding to the quantum number $l$ of the orbital angular momentum).
For the spinor at rest, $k=0$, so $j_0(kr)=\sin(kr)/kr\to 1$ and $j_1(kr)=\sin(kr)/(kr)^2-\cos(kr)/kr\to 0$.
In this case, the oscillatory spherical-wave structure flattens out and delocalizes; the wave function becomes uniform in all space.
}\\

{\bf 11. Charge conjugation and matter--antimatter asymmetry}\\ \\
The Dirac equation in spacetime with curvature and torsion is given by (\ref{Diracfinal}):
\begin{equation}
    i\hbar \gamma^\mu e_\mu^i\mathring{\nabla}_i\psi = mc\psi - \mathcal{K}(\overline{\psi}\gamma^\mu\gamma^5\psi)\gamma_\mu\gamma^5\psi.
    \label{Diractorsion}
\end{equation}
The charge conjugate $\psi_c$ of a spinor $\psi$ is defined as \cite{RQM,QFT}
\begin{equation}
    \psi_c=i\gamma^2\psi^\ast,\quad \psi^\ast=i\gamma^2\psi_c.
    \label{conjugate}
\end{equation}
The double charge-conjugation transformation is equivalent to the identity transformation: $(\psi_c)_c=i\gamma^2(\psi_c)^\ast=\psi$.
Using $\gamma^{\mu\dagger}=\gamma^0\gamma^\mu\gamma^0$, $\gamma^{5\dagger}=\gamma^5$, and $\{\gamma^\mu,\gamma^5\}=0$ gives the reality of the axial vector $\overline{\psi}\gamma^\mu\gamma^5\psi$:
\begin{equation}
    (\overline{\psi}\gamma^\mu\gamma^5\psi)^\ast=(\psi^\dagger\gamma^0\gamma^\mu\gamma^5\psi)^\dagger=\psi^\dagger\gamma^{5\dagger}\gamma^{\mu\dagger}\gamma^{0\dagger}\psi=\overline{\psi}\gamma^\mu\gamma^5\psi.
    \label{real}
\end{equation}
For a spinor with electric charge $q$ in the presence of the electromagnetic potential $A_i$, the covariant derivative $\mathring{\nabla}_i\psi$ is generalized to $\mathring{\nabla}_i\psi-i(q/\hbar c)A_i\psi$.
The nonlinear Dirac equation (\ref{Diractorsion}) is then generalized to
\begin{equation}
    i\hbar \gamma^\mu e_\mu^i\mathring{\nabla}_i\psi + \frac{q}{c}e^i_\mu A_i\gamma^\mu\psi = mc\psi - \mathcal{K}(\overline{\psi}\gamma^\mu\gamma^5\psi)\gamma_\mu\gamma^5\psi.
    \label{Diracpotential}
\end{equation}

The complex conjugate of (\ref{Diracpotential}) is, using $\gamma^{5\ast}=\gamma^5$:
\[
    -i\hbar \gamma^{\mu\ast} e_\mu^i\mathring{\nabla}_i\psi^\ast + \frac{q}{c}e^i_\mu A_i\gamma^{\mu\ast}\psi^\ast = mc\psi^\ast - \mathcal{K}(\overline{\psi}\gamma^\mu\gamma^5\psi)\gamma_\mu^\ast\gamma^5\psi^\ast.
\]
Putting (\ref{conjugate}) and $\gamma^{\mu\ast}=\gamma^2\gamma^\mu\gamma^2$ in this relation gives
\[
    -i\hbar e^i_\mu\gamma^2\gamma^\mu\gamma^2(i\gamma^2)\mathring{\nabla}_i\psi_c+\frac{q}{c}e^i_\mu A_i\gamma^2\gamma^\mu\gamma^2(i\gamma^2)\psi_c=mc(i\gamma^2)\psi_c-\mathcal{K}(\overline{\psi}\gamma^\mu\gamma^5\psi)\gamma^2\gamma_\mu\gamma^2\gamma^5(i\gamma^2)\psi_c,
\]
which is equivalent to
\[
    \hbar e^i_\mu\gamma^2\gamma^\mu\mathring{\nabla}_i\psi_c+i\frac{q}{c}e^i_\mu A_i\gamma^2\gamma^\mu\psi_c=-imc\gamma^2\psi_c+i\mathcal{K}(\overline{\psi}\gamma^\mu\gamma^5\psi)\gamma^2\gamma_\mu\gamma^5\psi_c.
\]
Multiplying this equation by $-i\gamma^2$ from the left brings it to the form analogous to (\ref{Diracpotential}):
\begin{equation}
    i\hbar e^i_\mu\gamma^\mu\mathring{\nabla}_i\psi_c-\frac{q}{c}e^i_\mu A_i\gamma^\mu\psi_c=mc\psi_c-\mathcal{K}(\overline{\psi}\gamma^\mu\gamma^5\psi)\gamma_\mu\gamma^5\psi_c.
    \label{Diracconjugated}
\end{equation}

The product $(\overline{\psi}\gamma^\mu\gamma^5\psi)$ in (\ref{Diracconjugated}) must be written in terms of $\psi_c$.
The Hermitian conjugate of (\ref{conjugate}) gives
\[
    \psi^T=\psi^{\ast\dagger}=-i\psi_c^\dagger\gamma^{2\dagger}=i\psi_c^\dagger\gamma^2.
\]
Therefore, using the reality relation (\ref{real}), this product is
\begin{eqnarray}
    & & \overline{\psi}\gamma^\mu\gamma^5\psi=(\psi^\dagger\gamma^0\gamma^\mu\gamma^5\psi)^\ast=\psi^T(\gamma^2\gamma^0\gamma^2)(\gamma^2\gamma^\mu\gamma^2)\gamma^5\psi^\ast=(i\psi_c^\dagger\gamma^2)(\gamma^2\gamma^0\gamma^2)(\gamma^2\gamma^\mu\gamma^2)\gamma^5(i\gamma^2\psi_c) \nonumber \\
    & & =-\psi_c^\dagger\gamma^0\gamma^\mu\gamma^2\gamma^5\gamma^2\psi_c=-\overline{\psi_c}\gamma^\mu\gamma^5\psi_c. \nonumber
\end{eqnarray}
Putting this relation in (\ref{Diracconjugated}) gives the Dirac equation with torsion for the charge-conjugated spinor field $\psi_c$ \cite{anti}:
\begin{equation}
    i\hbar e^i_\mu\gamma^\mu\mathring{\nabla}_i\psi_c-\frac{q}{c}e^i_\mu A_i\gamma^\mu\psi_c=mc\psi_c+\mathcal{K}(\overline{\psi_c}\gamma^\mu\gamma^5\psi_c)\gamma_\mu\gamma^5\psi_c.
    \label{Diracantimatter}
\end{equation}

Comparing (\ref{Diracpotential}) with (\ref{Diracantimatter}) shows that $\psi$ and $\psi_c$ correspond to the opposite values of $q$: the charge-conjugation transformation changes the sign of the electric charge of a spinor \cite{RQM,QFT}.
However, the field equations for $\psi$ and $\psi_c$ in spacetime with torsion are different because of the opposite signs of the cubic term relative to the mass term \cite{anti}.
This asymmetry is related to that the scalar Dirac bilinear $\overline{\psi}\psi$ changes sign under the charge-conjugation transformation: $\overline{\psi_c}\psi_c=-\overline{\psi}\psi$, whereas the square of the axial current $\overline{\psi}\gamma^\mu\gamma^5\psi$ does not change sign: $(\overline{\psi_c}\gamma^\mu\gamma^5\psi_c)(\overline{\psi_c}\gamma_\mu\gamma^5\psi_c)=(\overline{\psi}\gamma^\mu\gamma^5\psi)(\overline{\psi}\gamma_\mu\gamma^5\psi)$.

This asymmetry can be shown for the spherical spinors (\ref{spinorsolution}).
The charge conjugate $\xi_c$ of a Pauli spinor $\xi$ is
\[
    \xi_c=i\sigma_2\xi^\ast,
\]
analogously to (\ref{conjugate}).
The charge conjugate of the negative-energy state $\tilde{\psi}^{-}$ is therefore
\begin{eqnarray}
    & & \tilde{\psi}^{-}_c=i\gamma^2\tilde{\psi}^{-\ast}=\left( \begin{array}{cc}
    0 & i\sigma_2 \\
    -i\sigma_2 & 0 \end{array} \right)\left( \begin{array}{c}
    -if({\bm\sigma}^\ast\cdot{\bf n})\xi^\ast \\
    g\xi^\ast \\ \end{array} \right)=\left( \begin{array}{c}
    ig\sigma_2\xi^\ast \\
    -f\sigma_2({\bm\sigma}^\ast\cdot{\bf n})\xi^\ast \\ \end{array} \right)=\left( \begin{array}{c}
    ig\sigma_2\xi^\ast \\
    f({\bm\sigma}\cdot{\bf n})\sigma_2\xi^\ast \\ \end{array} \right) \nonumber \\
    & & =\left( \begin{array}{c}
    g\xi_c \\
    -if({\bm\sigma}\cdot{\bf n})\xi_c \\ \end{array} \right),
\end{eqnarray}
using the identity $\sigma_2{\bm\sigma}^\ast=-{\bm\sigma}\sigma_2$, equivalent to ${\bm\sigma}^\ast=-\sigma_2{\bm\sigma}\sigma_2$.
This conjugate is equal to the positive-energy state $\tilde{\psi}^{+}$ constructed from $\xi_c$ instead of $\xi$, with the function $f(r)$ replaced with $-f(r)$, and with the energy $E$ replaced with $-E$ (which corresponds to the time-dependence factor $e^{iEt/\hbar}$ being conjugated to $e^{-iEt/\hbar}$).
Similarly, the charge conjugate of the positive-energy state $\tilde{\psi}^{+}$ is
\begin{equation}
    \tilde{\psi}^{+}_c=i\gamma^2\tilde{\psi}^{+\ast}=\left( \begin{array}{c}
    if({\bm\sigma}\cdot{\bf n})\xi_c \\
    -g\xi_c \\ \end{array} \right).
\end{equation}
This conjugate is equal to the negative-energy state $\tilde{\psi}^{-}$ constructed from $\xi_c$ instead of $\xi$, with the function $g(r)$ replaced with $-g(r)$ (equivalent to $f(r)$ replaced with $-f(r)$ up to an overall sign change), and with the energy $E$ replaced with $-E$.

Changing $f$ to $-f$ and $E$ to $-E$ in the upper-component negative-energy equation (\ref{uppernegative}) brings it to the lower-component positive-energy equation (\ref{lowerpositive}) with the constant $\mathcal{K}$ changed to $-\mathcal{K}$.
Changing $f$ to $-f$ and $E$ to $-E$ in the lower-component negative-energy equation (\ref{lowernegative}) brings it to the upper-component positive-energy equation (\ref{upperpositive}) with the constant $\mathcal{K}$ changed to $-\mathcal{K}$.
This asymmetry corresponds to the sign difference of the cubic term relative to the mass term in the Dirac equations 
(\ref{Diracpotential}) and (\ref{Diracantimatter}).
A positive-energy state (representing a particle of matter) and a charge-conjugated negative-energy state (representing the corresponding particle of antimatter) in spacetime with torsion satisfy different field equations.
This difference may explain the observed matter--antimatter asymmetry in particle physics.\\

{\bf 12. The energy--momentum tensor for a spherical spinor with curvature and torsion}\\ \\
The combined energy--momentum tensor for a spherical Dirac spinor in spacetime with curvature and torsion is given by (\ref{combinedDirac}), in which $\psi$ is changed to $\Psi$ and $\gamma^\mu$ is changed to $\tilde{\gamma}^\mu$:
\begin{equation}
    T_{ik}+U_{ik}=\frac{1}{4}i\hbar c\left[\overline{\Psi}\tilde{\gamma}_\mu(e^\mu_i\mathring{\nabla}_k\Psi+e^\mu_k\mathring{\nabla}_i\Psi)-(e^\mu_i\mathring{\nabla}_k\overline{\Psi}+e^\mu_k\mathring{\nabla}_i\overline{\Psi})\tilde{\gamma}_\mu\Psi\right]+\frac{1}{2}\mathcal{K}c j^{5\mu} j^5_\mu g_{ik},
    \label{combinedrotated}
\end{equation}
where $\Psi$ denotes the rotated spinor:
\[
\Psi=\tilde{\psi}.
\]
This rotation cancels the spinor connections $\Gamma_\theta$ and $\Gamma_\phi$ without changing $T_{ik}$, according to the results of Section 8.
The tensor $T_{ik}+U_{ik}$ does not explicitly depend on the spinor mass; this dependence appears through functions $g(r)$ and $f(r)$, which are solutions of the Dirac equation containing $m$.

For the positive-energy spinor $\Psi^{+}$ in (\ref{spinorsolution}), the term $j^{5\mu} j^5_\mu$ is determined by (\ref{axialpositive}):
\[
    j^{5\mu} j^5_\mu=[g\xi^\dagger\quad if\xi^\dagger({\bm\sigma}\cdot{\bf n})]\left[-(g^2-f^2)\left( \begin{array}{c}
    g\xi \\
    if({\bm\sigma}\cdot{\bf n})\xi \\ \end{array} \right)+2igf({\bf s}\cdot{\bf n})\left( \begin{array}{c}
    if({\bm\sigma}\cdot{\bf n})\xi \\ g\xi \end{array} \right)\right]=-(g^2-f^2)^2-\frac{4}{3}(gf)^2,
\]
following the calculations in Section 10.
The last term in (\ref{combinedrotated}) is therefore equal to
\[
    -\frac{1}{2}\mathcal{K}c\left((g^2-f^2)^2+\frac{4}{3}(gf)^2\right),
\]
giving a negative contribution from torsion to the spinor energy.

For the negative-energy spinor $\Psi^{-}$ in (\ref{spinorsolution}), the term $j^{5\mu} j^5_\mu$ is determined by (\ref{axialnegative}) and has the same value.
However, for the charge-conjugated negative-energy spinor that represents a particle with positive energy, the constant $\mathcal{K}$ in the Dirac equation is replaced with $-\mathcal{K}$, according to the results of Section 11, giving a positive contribution from torsion to the spinor energy.
Consequently, the spin-torsion interaction in the Einstein--Cartan theory of gravity generates the matter--antimatter asymmetry: a free particle of antimatter (charge-conjugated negative-energy spinor) at rest has a higher energy than the corresponding free particle of matter (positive-energy spinor) at rest.

The component $T_{tt}+U_{tt}$ of the combined energy--momentum tensor (\ref{combinedrotated}) for the positive-energy spinor is
\begin{eqnarray}
    & & T_{tt}+U_{tt}=\frac{1}{2}i\hbar c\left[\overline{\Psi}\gamma_0e^0_t\mathring{\nabla}_{ct}\Psi-e^0_t\mathring{\nabla}_{ct}\overline{\Psi}\gamma_0\Psi\right]+\frac{1}{2}\mathcal{K}c j^{5\mu} j^5_\mu g_{tt} \nonumber \\
    & & =\frac{1}{2}i\hbar c\left[\overline{\Psi}\gamma_0e^0_t\partial_{ct}\Psi-e^0_t\partial_{ct}\overline{\Psi}\gamma_0\Psi-\overline{\Psi}\gamma_0e^0_t\tilde{\Gamma}_t\Psi-e^0_t\overline{\Psi}\tilde{\Gamma}_t\gamma_0\Psi\right]+\frac{1}{2}\mathcal{K}c j^{5\mu} j^5_\mu g_{tt} \nonumber \\
    & & =Ee^{\nu/2}\overline{\Psi}\gamma^0\Psi+\frac{1}{2}\mathcal{K}c j^{5\mu} j^5_\mu e^\nu=Ee^{\nu/2}\Psi^\dagger\Psi+\frac{1}{2}\mathcal{K}c j^{5\mu} j^5_\mu e^\nu \nonumber \\
    & & =Ee^{\nu/2}(g^2+f^2)-\frac{1}{2}\mathcal{K}c e^\nu\left((g^2-f^2)^2+\frac{4}{3}(gf)^2\right), \nonumber
\end{eqnarray}
where the spinor connections are canceled because $\tilde{\Gamma}_t$ is proportional to $\gamma^0\tilde{\gamma}^3$ (\ref{spinorconnection}), which anticommutes with $\gamma^0$.
The corresponding mixed component is therefore
\begin{equation}
    T_t^t+U_t^t=Ee^{-\nu/2}(g^2+f^2)-\frac{1}{2}\mathcal{K}c\left((g^2-f^2)^2+\frac{4}{3}(gf)^2\right),
    \label{energypositive}
\end{equation}
which is equal to the energy density $\epsilon$.
For the charge-conjugated negative-energy spinor, the energy density is
\begin{equation}
    T_t^t+U_t^t=|E|e^{-\nu/2}(g^2+f^2)+\frac{1}{2}\mathcal{K}c\left((g^2-f^2)^2+\frac{4}{3}(gf)^2\right).
    \label{energynegative}
\end{equation}

The quantity $E$ appearing in (\ref{energypositive}) and (\ref{energynegative}) corresponds to the oscillating phase of the spinor: $\Psi^{+}\sim e^{-iEt/\hbar}$ and $\Psi^{-}\sim e^{iEt/\hbar}$.
The actual energy of the spinor is determined by the volume integral:
\begin{equation}
    \mathfrak{E}=\int\epsilon\sqrt{\mathfrak{s}}dV=4\pi\int_0^\infty(T_t^t+U_t^t) e^{\lambda/2} r^2 dr.
    \label{actualenergy}
\end{equation}
where $\mathfrak{s}$ is the determinant of the spatial metric tensor $\gamma_{ik}=-g_{ik}+g_{0i}g_{ok}/g_{00}$ ($0$ being the coordinate $ct$), which for the metric (\ref{centralmetric}) is $\mathfrak{s}=e^\lambda r^4\sin^2\theta$.
The ground state of the spinor is determined by the lowest minimum of the function $\mathfrak{E}(E)$ (\ref{actualenergy}) with respect to the parameter $E$: $\partial\mathfrak{E}/\partial E=0$ \cite{Soler}.
The ground state of the charge-conjugated negative-energy spinor has a higher energy than the ground state of the corresponding positive-energy spinor.

We calculate the other components of the tensor (\ref{combinedrotated}) for the positive-energy spinor $\Psi^{+}$.
The component $T_{rr}+U_{rr}$ is
\[
    T_{rr}+U_{rr}=\frac{1}{2}i\hbar c[\overline{\Psi}\tilde{\gamma}_3 e^3_r\partial_r\Psi-e^3_r\partial_r\overline{\Psi}\tilde{\gamma}_3\Psi]+\frac{1}{2}\mathcal{K}c j^{5\mu} j^5_\mu g_{rr}=-\frac{1}{2}i\hbar ce^3_r[\overline{\Psi}\gamma^r\partial_r\Psi-\partial_r\overline{\Psi}\gamma^r\Psi]+\frac{1}{2}\mathcal{K}c j^{5\mu} j^5_\mu g_{rr}.
\]
The terms with $\partial_r$ give
\[
    \overline{\Psi}\gamma^r\partial_r\Psi-\partial_r\overline{\Psi}\gamma^r\Psi=[g\xi^\dagger\quad if\xi^\dagger\sigma_r]\left( \begin{array}{cc}
    0 & \sigma_r \\
    -\sigma_r & 0 \end{array} \right)\left( \begin{array}{c}
    g'\xi \\
    if'\sigma_r\xi \\ \end{array} \right)-[g'\xi^\dagger\quad if'\xi^\dagger\sigma_r]\left( \begin{array}{cc}
    0 & \sigma_r \\
    -\sigma_r & 0 \end{array} \right)\left( \begin{array}{c}
    g\xi \\
    if\sigma_r\xi \\ \end{array} \right)=2i(f'g-g'f),
\]
so
\[
    T_{rr}+U_{rr}=\hbar ce^{\lambda/2}(f'g-g'f)-\frac{1}{2}\mathcal{K}c j^{5\mu} j^5_\mu e^{\lambda}.
\]
The corresponding mixed component is therefore
\begin{equation}
    T_r^r+U_r^r=\hbar ce^{-\lambda/2}(g'f-f'g)-\frac{1}{2}\mathcal{K}c\left((g^2-f^2)^2+\frac{4}{3}(gf)^2\right).
    \label{radialpressure}
\end{equation}

The component $T_{\theta\theta}+U_{\theta\theta}$ is
\[
    T_{\theta\theta}+U_{\theta\theta}=\frac{1}{2}i\hbar c[\overline{\Psi}\tilde{\gamma}_1 e^1_\theta\partial_\theta\Psi-e^1_\theta\partial_\theta\overline{\Psi}\tilde{\gamma}_1\Psi]+\frac{1}{2}\mathcal{K}c j^{5\mu} j^5_\mu g_{\theta\theta}=-\frac{1}{2}i\hbar ce^1_\theta[\overline{\Psi}\gamma^\theta\partial_\theta\Psi-\partial_\theta\overline{\Psi}\gamma^\theta\Psi]+\frac{1}{2}\mathcal{K}c j^{5\mu} j^5_\mu g_{\theta\theta}.
\]
Using $\partial_\theta\sigma_r=\sigma_\theta$ (\ref{derivatives}), the terms with $\partial_\theta$ give
\[
    \overline{\Psi}\gamma^\theta\partial_\theta\Psi-\partial_\theta\overline{\Psi}\gamma^\theta\Psi=[g\xi^\dagger\quad if\xi^\dagger\sigma_r]\left( \begin{array}{cc}
    0 & \sigma_\theta \\
    -\sigma_\theta & 0 \end{array} \right)\left( \begin{array}{c}
    0 \\
    if\sigma_\theta\xi \\ \end{array} \right)-[0\quad if\xi^\dagger\sigma_\theta]\left( \begin{array}{cc}
    0 & \sigma_\theta \\
    -\sigma_\theta & 0 \end{array} \right)\left( \begin{array}{c}
    g\xi \\
    if\sigma_r\xi \\ \end{array} \right)=2igf,
\]
so
\[
    T_{\theta\theta}+U_{\theta\theta}=\hbar crgf-\frac{1}{2}\mathcal{K}c j^{5\mu} j^5_\mu r^2.
\]
The corresponding mixed component is therefore
\begin{equation}
    T_\theta^\theta+U_\theta^\theta=-\hbar cgf/r-\frac{1}{2}\mathcal{K}c\left((g^2-f^2)^2+\frac{4}{3}(gf)^2\right).
    \label{tangentialpressure1}
\end{equation}

The component $T_{\phi\phi}+U_{\phi\phi}$ is
\[
    T_{\phi\phi}+U_{\phi\phi}=\frac{1}{2}i\hbar c[\overline{\Psi}\tilde{\gamma}_2 e^2_\phi\partial_\phi\Psi-e^2_\phi\partial_\phi\overline{\Psi}\tilde{\gamma}_2\Psi]+\frac{1}{2}\mathcal{K}c j^{5\mu} j^5_\mu g_{\phi\phi}=-\frac{1}{2}i\hbar ce^2_\phi[\overline{\Psi}\gamma^\phi\partial_\phi\Psi-\partial_\phi\overline{\Psi}\gamma^\phi\Psi]+\frac{1}{2}\mathcal{K}c j^{5\mu} j^5_\mu g_{\phi\phi}.
\]
Using $\partial_\phi\sigma_r=\sin\theta\,\sigma_\phi$ (\ref{derivatives}), the terms with $\partial_\phi$ give
\[
    T_{\phi\phi}+U_{\phi\phi}=\hbar cr\sin^2\theta\,gf-\frac{1}{2}\mathcal{K}c j^{5\mu} j^5_\mu r^2\sin^2\theta.
\]
The corresponding mixed component is therefore
\begin{equation}
    T_\phi^\phi+U_\phi^\phi=-\hbar cgf/r-\frac{1}{2}\mathcal{K}c\left((g^2-f^2)^2+\frac{4}{3}(gf)^2\right),
    \label{tangentialpressure2}
\end{equation}
which is equal to (\ref{tangentialpressure1}) in accordance with the central symmetry of the spherical spinor.

The component $T_{tr}+U_{tr}$ is
\begin{eqnarray}
    & & T_{tr}+U_{tr}=\frac{1}{4}i\hbar c[\overline{\Psi}\gamma_0 e^0_t\partial_r\Psi+\overline{\Psi}\tilde{\gamma}_3 e^3_r\nabla_{ct}\Psi-e^0_t\partial_r\overline{\Psi}\gamma_0\Psi-e^3_r\nabla_{ct}\overline{\Psi}\tilde{\gamma}_3\Psi] \nonumber \\
    & & = \frac{1}{4}i\hbar c[\overline{\Psi}\gamma_0 e^0_t\partial_r\Psi+\overline{\Psi}\tilde{\gamma}_3 e^3_r\partial_{ct}\Psi-e^0_t\partial_r\overline{\Psi}\gamma_0\Psi-e^3_r\partial_{ct}\overline{\Psi}\tilde{\gamma}_3\Psi], \nonumber \\
    & & =\frac{1}{4}i\hbar c e^0_t[\Psi^\dagger\partial_r\Psi-\partial_r\Psi^\dagger\Psi]-\frac{1}{2}Ee^3_r\overline{\Psi}\gamma^r\Psi, \nonumber
\end{eqnarray}
where spinor connections are canceled because $\tilde{\Gamma}_t$ anticommutes with $\tilde{\gamma}^3$.
The terms with $\partial_r$ give
\[
    \Psi^\dagger\partial_r\Psi-\partial_r\Psi^\dagger\Psi=[g\xi^\dagger\quad -if\xi^\dagger\sigma_r]\left( \begin{array}{c}
    g'\xi \\
    if'\sigma_r\xi \\ \end{array} \right)-[g'\xi^\dagger\quad -if'\xi^\dagger\sigma_r]\left( \begin{array}{c}
    g\xi \\
    if\sigma_r\xi \\ \end{array} \right)=0,
\]
and the term with $E$ gives
\[
    \overline{\Psi}\gamma^r\Psi=[g\xi^\dagger\quad if\xi^\dagger\sigma_r]\left( \begin{array}{cc}
    0 & \sigma_r \\
    -\sigma_r & 0 \end{array} \right)\left( \begin{array}{c}
    g\xi \\
    if\sigma_r\xi \\ \end{array} \right)=0.
\]
Consequently, $T_t^r+U_t^r=0$.

The component $T_{r\theta}+U_{r\theta}$ is
\begin{eqnarray}
    & & T_{r\theta}+U_{r\theta}=\frac{1}{4}i\hbar c[\overline{\Psi}\tilde{\gamma}_3 e^3_r\partial_\theta\Psi+\overline{\Psi}\tilde{\gamma}_1 e^1_\theta\partial_r\Psi-e^3_r\partial_\theta\overline{\Psi}\tilde{\gamma}_3\Psi-e^1_\theta\partial_r\overline{\Psi}\tilde{\gamma}_1\Psi] \nonumber \\
    & & =-\frac{1}{4}i\hbar c[\overline{\Psi}\gamma^r e^3_r\partial_\theta\Psi+\overline{\Psi}\gamma^\theta e^1_\theta\partial_r\Psi-e^3_r\partial_\theta\overline{\Psi}\gamma^r\Psi-e^1_\theta\partial_r\overline{\Psi}\gamma^\theta\Psi]. \nonumber
\end{eqnarray}
The terms with $\partial_r$ give
\[
    \overline{\Psi}\gamma^\theta\partial_r\Psi-\partial_r\overline{\Psi}\gamma^\theta\Psi=[g\xi^\dagger\quad if\xi^\dagger\sigma_r]\left( \begin{array}{cc}
    0 & \sigma_\theta \\
    -\sigma_\theta & 0 \end{array} \right)\left( \begin{array}{c}
    g'\xi \\
    if'\sigma_r\xi \\ \end{array} \right)-[g'\xi^\dagger\quad if'\xi^\dagger\sigma_r]\left( \begin{array}{cc}
    0 & \sigma_\theta \\
    -\sigma_\theta & 0 \end{array} \right)\left( \begin{array}{c}
    g\xi \\
    if\sigma_r\xi \\ \end{array} \right)=0,
\]
and the terms with $\partial_\theta$ give
\[
    \overline{\Psi}\gamma^r\partial_\theta\Psi-\partial_\theta\overline{\Psi}\gamma^r\Psi=[g\xi^\dagger\quad if\xi^\dagger\sigma_r]\left( \begin{array}{cc}
    0 & \sigma_r \\
    -\sigma_r & 0 \end{array} \right)\left( \begin{array}{c}
    0 \\
    if\sigma_\theta\xi \\ \end{array} \right)-[0\quad if\xi^\dagger\sigma_\theta]\left( \begin{array}{cc}
    0 & \sigma_r \\
    -\sigma_r & 0 \end{array} \right)\left( \begin{array}{c}
    g\xi \\
    if\sigma_r\xi \\ \end{array} \right)=0.
\]
Consequently, $T_r^\theta+U_r^\theta=0$.
Analogously, the component $T_{r\phi}+U_{r\phi}$ is
\[
    T_{r\phi}+U_{r\phi}=\frac{1}{4}i\hbar c[\overline{\Psi}\tilde{\gamma}_3 e^3_r\partial_\phi\Psi+\overline{\Psi}\tilde{\gamma}_2 e^2_\phi\partial_r\Psi-e^3_r\partial_\phi\overline{\Psi}\tilde{\gamma}_3\Psi-e^2_\phi\partial_r\overline{\Psi}\tilde{\gamma}_2\Psi],
\]
giving $T_r^\phi+U_r^\phi=0$.

The component $T_{\theta\phi}+U_{\theta\phi}$ is
\begin{eqnarray}
    & & T_{\theta\phi}+U_{\theta\phi}=\frac{1}{4}i\hbar c[\overline{\Psi}\tilde{\gamma}_1 e^1_\theta\partial_\phi\Psi+\overline{\Psi}\tilde{\gamma}_2 e^2_\phi\partial_\theta\Psi-e^1_\theta\partial_\phi\overline{\Psi}\tilde{\gamma}_1\Psi-e^2_\phi\partial_\theta\overline{\Psi}\tilde{\gamma}_2\Psi] \nonumber \\
    & & =-\frac{1}{4}i\hbar c[\overline{\Psi}\gamma^\theta e^1_\theta\partial_\phi\Psi+\overline{\Psi}\gamma^\phi e^2_\phi\partial_\theta\Psi-e^1_\theta\partial_\phi\overline{\Psi}\gamma^\theta\Psi-e^2_\phi\partial_\theta\overline{\Psi}\gamma^\phi\Psi]. \nonumber
\end{eqnarray}
The terms with $\partial_\theta$ give
\[
    \overline{\Psi}\gamma^\phi\partial_\theta\Psi-\partial_\theta\overline{\Psi}\gamma^\phi\Psi=[g\xi^\dagger\quad if\xi^\dagger\sigma_r]\left( \begin{array}{cc}
    0 & \sigma_\phi \\
    -\sigma_\phi & 0 \end{array} \right)\left( \begin{array}{c}
    0 \\
    if\sigma_\theta\xi \\ \end{array} \right)-[0\quad if\xi^\dagger\sigma_\theta]\left( \begin{array}{cc}
    0 & \sigma_\phi \\
    -\sigma_\phi & 0 \end{array} \right)\left( \begin{array}{c}
    g\xi \\
    if\sigma_r\xi \\ \end{array} \right)=0,
\]
and the terms with $\partial_\phi$ give
\[
    \overline{\Psi}\gamma^\theta\partial_\phi\Psi-\partial_\phi\overline{\Psi}\gamma^\theta\Psi=[g\xi^\dagger\quad if\xi^\dagger\sigma_r]\left( \begin{array}{cc}
    0 & \sigma_\theta \\
    -\sigma_\theta & 0 \end{array} \right)\left( \begin{array}{c}
    0 \\
    if\sin\theta\,\sigma_\phi\xi \\ \end{array} \right)-[0\quad if\sin\theta\,\xi^\dagger\sigma_\phi]\left( \begin{array}{cc}
    0 & \sigma_\theta \\
    -\sigma_\theta & 0 \end{array} \right)\left( \begin{array}{c}
    g\xi \\
    if\sigma_r\xi \\ \end{array} \right)=0.
\]
Consequently, $T_\theta^\phi+U_\theta^\phi=0$.

The components
\[
    T_{t\theta}+U_{t\theta}=\frac{1}{4}i\hbar c[\overline{\Psi}\gamma_0 e^0_t\partial_\theta\Psi+\overline{\Psi}\tilde{\gamma}_1 e^1_\theta\nabla_{ct}\Psi-e^0_t\partial_\theta\overline{\Psi}\gamma_0\Psi-e^1_\theta\nabla_{ct}\overline{\Psi}\tilde{\gamma}_1\Psi]
\]
and
\[
    T_{t\phi}+U_{t\phi}=\frac{1}{4}i\hbar c[\overline{\Psi}\gamma_0 e^0_t\partial_\phi\Psi+\overline{\Psi}\tilde{\gamma}_2 e^2_\phi\nabla_{ct}\Psi-e^0_t\partial_\phi\overline{\Psi}\gamma_0\Psi-e^2_\phi\nabla_{ct}\overline{\Psi}\tilde{\gamma}_2\Psi]
\]
include spinor connections.
However, they can be calculated directly using the conservation of the energy--momentum tensor $\mathring{\nabla}_i(T^i_k+U^i_k)=0$, which follows from the temporal contracted Bianchi identity $\mathring{\nabla}_i G^i_k=0$, for the metric (\ref{centralmetric}):
\[
    \mathring{\nabla}_i (T+U)^i_t=\partial_i (T+U)^i_t-\mathring{\Gamma}^k_{ti}(T+U)^i_k+\mathring{\Gamma}^i_{ki}(T+U)^k_t=\mathring{\Gamma}^\phi_{\theta\phi}(T+U)^\theta_t=\cot\theta\,(T+U)^\theta_t=0.
\]
Consequently, $T_t^\theta+U_t^\theta=0$.
Because of the central symmetry, $T_t^\phi+U_t^\phi=0$.

All components of $T_i^k+U_i^k$ for the charge-conjugated negative-energy spinor $\Psi^-_c$ are obtained from the positive-energy spinor $\Psi^+$ by changing $\mathcal{K}$ to $-\mathcal{K}$,
according to the asymmetric sign difference of the cubic term relative to the mass term in the Dirac equations 
(\ref{Diracpotential}) and (\ref{Diracantimatter}).\\

{\bf 13. Einstein--Cartan--Dirac equations for a spherical spinor}\\ \\
In a static gravitational field, corresponding to a stationary spinor, the components of the energy--momentum tensor $T^i_k+U^i_k$ (\ref{combinedrotated}) determine the energy density and the pressure of the spinor.
The energy density is given by 
$T_t^t+U_t^t=\epsilon$ (\ref{energypositive}).
The component (\ref{radialpressure}) $T_r^r+U_r^r=-p_\textrm{rad}$, where $p_\textrm{rad}$ is the radial pressure.
The components (\ref{tangentialpressure1}) and (\ref{tangentialpressure2}) $T_\theta^\theta+U_\theta^\theta=T_\phi^\phi+U_\phi^\phi=-p_\textrm{tan}$, where $p_\textrm{tan}$ is the tangential pressure.
The equality of these components also follows from the polar contracted Bianchi identity $\mathring{\nabla}_i (T+U)^i_\theta=\partial_i (T+U)^i_\theta-\mathring{\Gamma}^k_{\theta i}(T+U)^i_k+\mathring{\Gamma}^i_{ki}(T+U)^k_\theta=\mathring{\Gamma}^\phi_{\theta\phi}[(T+U)^\theta_\theta-(T+U)^\phi_\phi]=\cot\theta[(T+U)^\theta_\theta-(T+U)^\phi_\phi]=0$.
The component $T_t^r+U_t^r=0$ determines the radial energy flux, proportional to the radial momentum density of a stationary spinor.
The components $T_t^\theta+U_t^\theta=0$ and $T_t^\phi+U_t^\phi=0$ determine the angular energy flux, proportional to the orbital angular momentum density of a stationary spinor.

The nonzero components of the Einstein tensor $G^i_k$ corresponding to the centrally symmetric metric (\ref{centralmetric}) are \cite{LL2}
\begin{eqnarray}
    & & G^t_t=\frac{1}{r^2}-e^{-\lambda}\left(\frac{1}{r^2}-\frac{\lambda'}{r}\right),\quad G^r_r=\frac{1}{r^2}-e^{-\lambda}\left(\frac{1}{r^2}+\frac{\nu'}{r}\right),\quad G^r_t=-e^{-\lambda}\frac{\dot{\lambda}}{r}, \nonumber \\
    & & G^\theta_\theta=G^\phi_\phi=-\frac{1}{2}e^{-\lambda}\left(\nu''+\frac{\nu'^2}{2}-\frac{\nu'\lambda'}{2}+\frac{\nu'-\lambda'}{r}\right)+\frac{1}{2}e^{-\nu}\left(\ddot{\lambda}+\frac{\dot{\lambda}^2}{2}-\frac{\dot{\nu}\dot{\lambda}}{2}\right). \nonumber
    \label{centralgravity}
\end{eqnarray}
In the static field, they reduce to
\begin{eqnarray}
    & & G^t_t=\frac{1}{r^2}-e^{-\lambda}\left(\frac{1}{r^2}-\frac{\lambda'}{r}\right),\quad G^r_r=\frac{1}{r^2}-e^{-\lambda}\left(\frac{1}{r^2}+\frac{\nu'}{r}\right),\quad G^r_t=0, \nonumber \\
    & & G^\theta_\theta=G^\phi_\phi=-\frac{1}{2}e^{-\lambda}\left(\nu''+\frac{\nu'^2}{2}-\frac{\nu'\lambda'}{2}+\frac{\nu'-\lambda'}{r}\right).
    \label{staticgravity}
\end{eqnarray}
The gravitational field is coupled to positive-energy spinors.
The Einstein equations (\ref{Einstein}) for the Einstein tensor (\ref{staticgravity}) and the energy--momentum tensor with components (\ref{energypositive}), (\ref{radialpressure}), (\ref{tangentialpressure1}), and (\ref{tangentialpressure2}) therefore give
\begin{eqnarray}
    & & \frac{1}{r^2}-e^{-\lambda}\left(\frac{1}{r^2}-\frac{\lambda'}{r}\right)=\kappa Ee^{-\nu/2}(g^2+f^2)-\frac{1}{2}\kappa\mathcal{K}c\left((g^2-f^2)^2+\frac{4}{3}(gf)^2\right), \label{ECD1} \\
    & & \frac{1}{r^2}-e^{-\lambda}\left(\frac{1}{r^2}+\frac{\nu'}{r}\right)=\kappa\hbar ce^{-\lambda/2}(g'f-f'g)-\frac{1}{2}\kappa\mathcal{K}c\left((g^2-f^2)^2+\frac{4}{3}(gf)^2\right), \label{ECD2} \\
    & & -\frac{1}{2}e^{-\lambda}\left(\nu''+\frac{\nu'^2}{2}-\frac{\nu'\lambda'}{2}+\frac{\nu'-\lambda'}{r}\right)=-\kappa\hbar cgf/r-\frac{1}{2}\kappa\mathcal{K}c\left((g^2-f^2)^2+\frac{4}{3}(gf)^2\right). \nonumber
\end{eqnarray}
The last equation is not independent; it follows from the radial contracted Bianchi identity, which relates $(T+U)^\theta_\theta$ to $(T+U)^t_t$, $(T+U)^r_r$, and $\partial_r(T+U)^r_r$:
\begin{eqnarray}
    & & \mathring{\nabla}_i (T+U)^i_r=\partial_i (T+U)^i_r-\mathring{\Gamma}^k_{ri}(T+U)^i_k+\mathring{\Gamma}^i_{ki}(T+U)^k_r \nonumber \\
    & & =\partial_r (T+U)^r_r-\mathring{\Gamma}^t_{rt}(T+U)^t_t-\mathring{\Gamma}^\theta_{r\theta}(T+U)^\theta_\theta-\mathring{\Gamma}^\phi_{r\phi}(T+U)^\phi_\phi+\mathring{\Gamma}^t_{rt}(T+U)^r_r+\mathring{\Gamma}^\theta_{r\theta}(T+U)T^r_r+\mathring{\Gamma}^\phi_{r\phi}(T+U)^r_r \nonumber \\
    & & =\partial_r(T+U)^r_r-\frac{1}{2}\nu'(T+U)^t_t-\frac{2}{r}(T+U)^\theta_\theta+\frac{1}{2}\nu'(T+U)^r_r+\frac{2}{r}(T+U)^r_r=0.
\end{eqnarray}
Consequently, a stationary spherical spinor (Wakano spinor), representing a particle of matter, and its centrally symmetric static gravitational field are determined by the Dirac equation for the upper and lower spinor components (\ref{upperpositive}) and (\ref{lowerpositive}), and the Einstein equations (\ref{ECD1}) and (\ref{ECD2}).
These four nonlinear differential equations determine the four unknown functions: $g(r)$, $f(r)$, $\nu(r)$, and $\lambda(r)$.
Without torsion, these equations reduce to the Einstein--Dirac equations for a static, spherically symmetric system \cite{Soler}.

The gravitational field is not coupled to negative-energy spinors, but to the corresponding charge-conjugated spinors, which are equivalent to positive-energy spinors \cite{curved}.
A stationary spherical spinor, representing a particle of antimatter, and its centrally symmetric static gravitational field are determined by these equations with $\mathcal{K}$ changed to $-\mathcal{K}$,
according to the asymmetric sign difference of the cubic term relative to the mass term in the Dirac equations 
(\ref{Diracpotential}) and (\ref{Diracantimatter}).

For a charged spinor, the derivatives $\partial_i$ in the Dirac equation are generalized to $\partial_i-i(q/\hbar c)A_i$, following (\ref{Diracpotential}).
The energy--momentum tensor $(T+U)_{ik}$ is generalized to $(T+U)_{ik}+\tilde{T}_{ik}$, where 
\[
    \tilde{T}_{ik}=\frac{1}{4\pi}\left(\frac{1}{4}g_{ik}F_{lm}F^{lm}-F_i^{\phantom{i}j}F_{kj}\right)
\]
is the energy--momentum tensor for the electromagnetic field and $F_{ik}=\mathring{\nabla}_i A_k-\mathring{\nabla}_k A_i$ is the electromagnetic field tensor, which satisfies the Maxwell equations $\mathring{\nabla}_i F^{ik}=(4\pi/c)j^k$ for the electromagnetic current $j^k=qc\overline{\psi}\gamma^k\psi$, with the Lorentz gauge $\mathring{\nabla}_i A^i=0$.

The curvature of spacetime acts as a gravitational binding mechanism, creating regular stable localized particle-like soliton solutions of the Einstein--Dirac equations and the Einstein--Dirac--Maxwell equations \cite{Smoller}.
At distances $r\gg\hbar/mc$, the gravitational field of a spinor has a form of the Schwarzschild metric.
For a spinor with electric charge, the gravitational field has a form of the Reissner--Nordstr\"{o}m metric.

At densities much smaller than the Cartan density \cite{non}, the only effects of torsion are in the self-interaction of the nonlinear spinor, and they are extremely small.
At these densities, the rest energies of corresponding matter and antimatter particles are almost equal, in accordance with the results of experimental measurements of the masses of the proton and antiproton \cite{antiproton}.
However, the matter--antimatter energy asymmetry is not zero and might be responsible for the observed spin asymmetry in the proton \cite{proton}.
This asymmetry could possibly be measured in experiments probing smaller distances and higher densities.

Because the effects of torsion (proportional to $\kappa^2$) are small compared to the effects of curvature (proportional to $\kappa$), a spherical spinor solution of the Einstein--Cartan--Dirac equations (\ref{upperpositive}), (\ref{lowerpositive}), (\ref{ECD1}), and (\ref{ECD2}) represents a regular localized standing solitary wave (particle-like soliton) in spacetime with curvature and torsion, like for spacetime without torsion \cite{Smoller}.
Because of torsion, solutions representing particles of matter and antimatter are slightly different.
In the nonrelativistic limit, the Einstein--Cartan--Dirac equations reduce to the Schr\"{o}dinger--Newton equation \cite{nonrelativistic}.
The ground state of a fermion, whose energy is given by the lowest minimum of the integral (\ref{actualenergy}), has a lower energy than the ground state of the corresponding antifermion.
This difference might have increased in the early Universe because of the background torsion, possibly creating a mechanism that has led to the matter--antimatter asymmetry in the current Universe.\\

I am grateful to Francisco Guedes and my Parents, Bo\.{z}enna Pop{\l}awska and Janusz Pop{\l}awski, for supporting this work.

\end{document}